**Advancing the Economic and Environmental Sustainability of Rare Earth Element Recovery from Phosphogypsum**

Authors: Adam Smerigan[a], Rui Shi*[ab]

[a]Department of Chemical Engineering, The Pennsylvania State University, University Park, PA, USA

[b]Institute of Energy and the Environment, The Pennsylvania State University, University Park, PA, USA

*rms6987@psu.edu

Keywords: rare earth elements (REE), technoeconomic analysis (TEA), life cycle assessment (LCA), global uncertainty and sensitivity analysis, green chemical process design

# Abstract

Transitioning to green energy technologies requires more sustainable and secure rare earth elements (REE) production. The current production of rare earth oxides (REOs) is completed by an energy and chemically intensive process from the mining of REE ores. Investigations into a more sustainable supply of REEs from secondary sources, such as toxic phosphogypsum (PG) waste, is vital to securing the REE supply chain. However, conventional solvent extraction to recover dilute REEs from PG waste is inefficient and has high environmental impact. In this work, we propose a treatment train for the recovery of REEs from PG which includes a bio-inspired adsorptive separation to generate a stream of pure REEs, and we assess its financial viability and environmental impacts under uncertainties through a "probabilistic sustainability" framework integrating life cycle assessment (LCA) and techno-economic analysis (TEA). Results show that in 87% of baseline scenario simulations, the internal rate of return (IRR) exceeded 15%, indicating that this system has the potential to be profitable. However, environmental impacts of the system are mixed. Specifically, the proposed system outperforms conventional systems (REO mining and PG stack treatment) in ecosystem quality and resource depletion, but has higher human health impacts. Scenario analysis shows that the system is profitable at capacities larger than 100,000 $kg \cdot h^{-1} \cdot$PG for PG with REE content above 0.5 wt%. The most dilute PG sources (0.02-0.1 wt% REE) are inaccessible using the current process scheme (limited by the cost of acid and subsequent neutralization) requiring further examination of new process schemes and improvements in technological performance. Overall, this study evaluates the sustainability of a first-of-its-kind REE recovery process from PG and uses these results to provide clear direction for advancing sustainable REE recovery from secondary sources.

# Broader Context

To meet climate goals using green energy technologies, a sustainable source of rare earth elements (REEs) must be secured. The current REE supply chain (dependent on unsustainable mining and solvent extraction practices in primarily one geographic location) is unable to meet the projected demand. To meet this demand, the investigation of more sustainable pathways utilizing secondary sources and new separation technologies will be necessary. However, progress is hindered by the lack of knowledge around the economic feasibility and environmental impact of REE recovery processes from secondary sources. One such source of REEs is phosphogypsum (PG), which is a toxic byproduct of fertilizer production (>300 million tonnes per year). PG is stored indefinitely in large lakes, or stacks, due to its classification as a technologically enhanced naturally occurring radioactive material. As these PG stacks grow, there is increasing concern over leaching and accidental discharges of this toxic heavy metal filled slurry into the environment. Here, we identify a potential pathway for REE recovery from phosphogypsum for waste remediation. This system's sustainability is evaluated using techno-economic analysis, life cycle assessment, and global uncertainty/sensitivity analysis. The knowledge and research direction identified here will advance REE recovery from secondary sources for green energy applications.

# 1. Introduction

A major energy transition to clean energy is required to meet climate goals (< 2°C increase in global temperatures).[1] The clean energy transition will require large volumes of rare earth elements (REEs) for use in wind turbines and permanent magnets in electric vehicles. Therefore, clean energy technologies will comprise approximately 40% of the demand for REEs in the future.[2] To meet this additional demand, a more stable and sustainable supply of REEs must be established.

Currently, approximately 90% of REE production occurs in China by mining REE containing ores (e.g., monazite, bastnaesite, and xenotime).[2] Conventional REE production uses hydrometallurgical pathways to convert ores into saleable rare earth oxides (REO).[3,4] The process begins with mining the ores followed by beneficiation and leaching to extract the REEs from the other components. This concentrated REE stream is then separated by up to 300 stages of solvent extraction[5] and refined into the final REO product which is sold to make other high value products (e.g., magnets). While this scheme has been successful and profitable, it consumes large volumes of chemicals due to the intense acid/base leaching and the inefficient solvent extraction separation. This chemical consumption leads to negative environmental impacts from the production and transport of chemicals and the long-term storage of acidic tailings[6,7] (which consume lots of land and can fail catastrophically).[2] The consumption of organic solvent and acid regenerant used in the solvent extraction has been shown to contribute up to 30% of the overall environmental impact of conventional REE production.[5] This result motivates the development of new separation processes that use less energy and fewer toxic chemicals. One alternative technology that has been developed is solid-phase adsorption which uses ion-exchange or chelating resins to adsorb REEs from solution. Adsorption can be highly selective at low REE concentrations (such as those in dilute secondary sources) and has reduced chemical consumption compared to solvent extraction. However, resins are prohibitively expensive for their capacity and have low production rates compared to solvent extraction.[8,9] New bio-based adsorbents that utilize proteins or peptides with high affinity and high separation factors between REEs could make solid-phase adsorption an attractive option for dilute REE recovery.[10,11] However, it is unclear whether this technology is economically feasible and more environmentally friendly than conventional techniques when implemented in systems for REE recovery.

Globally available secondary sources (e.g., product recycling[12–14], coal fly ash[15], acid mine tailings[16]) are potentially more sustainable alternatives to satisfy increasing REO demand.[17,18] Phosphogypsum (PG) is a potential secondary source that is a waste from fertilizer production (specifically phosphoric acid production). Currently, this PG waste is stored in large above ground stacks due to its classification as a toxic and radioactive waste.[19] Within Florida alone, there are 200 million tons of PG stored in stacks which amounts to 1,000 tons of REEs.[20] Considering the PG production rate in the U.S. (30 million tons PG·year$^{-1}$)[20], the annual consumption of REEs within the United States (9,000 tons REE·year$^{-1}$)[21] could be satisfied from PG alone assuming a minimum REE concentration of 0.02-0.03 wt% in PG.[22,23] Even higher PG REE concentrations (0.03-0.9 wt%) can be found around the world (e.g., Poland, Brazil, Russia)[22], which may lead to more profitable operations. Alternative PG remediation systems (e.g., road construction, brick production, and soil amendment) and other broader applications show some promise but require further evaluation to ensure they meet health and environmental standards. Additionally, these alternative systems cost up to 77 times more than conventional stack treatment.[24] Therefore, the extraction of REEs from PG should be explored as a potentially profitable and environmentally friendly alternative to stack treatment. One study of a pilot-scale system for REE recovery from PG in Poland showed profitability at high risk and higher environmental impact than the conventional stack treatment.[25] However, this study did not allocate any impacts to coproducts from the leaching operation. Further, this study had a limited scope considering only a mixed REE product (lower value compared to pure individual REOs after separation) and heavy reliance on fossil energy. Therefore, further work must rigorously explore process alternatives, especially in leaching, for sustainable REE recovery systems from PG.

Since the optimal process scheme for REE recovery from PG is still unknown, studies have primarily examined the removal of REEs from the PG crystal lattice. Many studies have accomplished this using recrystallization, carbonation, bioleaching, organic leaching, and most commonly, inorganic acid leaching.[26–30] These acid leaching experiments examined the effect of different parameters (e.g., leaching time, solvent ratio, temperature) and lixiviant (e.g., nitric acid, sulfuric acid, hydrochloric acid) on leaching efficiency. An advantage of using sulfuric acid is the co-production of gypsum from the leaching process, but sulfuric acid has a lower REE leaching efficiency (approximately 40%) compared to hydrochloric acid and nitric acid (approximately 60%).[22,31,32] Presently, very little work has been done that applies this leaching data, in combination with a suitable selective separation, to evaluate the potential profitability and environmental impact of a REE recovery system. Part of the reason for this lack of research may be the challenge of working with lab-scale technologies where missing data requires the use of assumptions and heuristics. However, work at this stage is especially rewarding due to the ease of making large process changes at this stage (low financial investment) and the ability to provide direction for future research for faster implementation.[33,34]

In this work, we developed a novel system for REE recovery from PG (abbreviated herein as the REEPS system). This system utilizes a bio-based selective separation to create a pure individual REO product while also remediating the PG waste. We performed techno-economic analysis (TEA) and life cycle assessment (LCA) to evaluate system economic feasibility and environmental impacts. To identify the most influential parameters on system sustainability, we completed a global uncertainty and sensitivity analysis. Further, we determined key research directions and technology alternatives by identifying process hotspots. Different scenarios for system capacity and PG REE content were evaluated to identify profitable conditions and clarify the limitations of the process scheme. Overall, this work establishes a foundation for sustainability assessment of REE recovery from PG, setting standards for evaluating the feasibility of novel REE recovery systems, and highlights the importances of sustainability assessment for processes with low technological readiness. The approaches and framework developed in this work can be utilized by the broader community for REE recovery efforts.

# 2. Methods

## 2.1. System overview

The system developed here, the recovery of rare earth elements from PG system (REEPS), achieves two objectives: 1) to remediate PG waste 2) to produce high purity (>99%) REOs. Figure 1a shows the system diagram used for the LCA broken down by process section including input and output flows. The complete process flow diagram (Figure 1b) shows each unit operation, input and output stream, and the phase of each stream. Each process section is organized by the same color shown in Figure 1a. The first section (leaching) takes the PG waste feedstock and extracts the REEs from the solid gypsum lattice to the liquid phase. Next, these REEs are separated from the leached solids and are concentrated using precipitation by oxalic acid. The mixed REE-oxalate precipitate is filtered and resuspended prior to the REE-specific bioadsorption (selective separation). The selective separation uses an REE selective biomolecule (e.g., Lanmodulin protein) attached to an agarose resin to create pure individual REE streams. In the refining section, each individual REE stream from the selective separation is then precipitated by oxalic acid, filtered, and calcined to the final REO product. The wastewater from the process is neutralized using sodium hydroxide and all heavy metals are precipitated after the addition of sodium phosphate to achieve high pH (around 8). The wastewater after this neutralization is sent to an onsite wastewater treatment facility (primary, secondary, and tertiary treatment) to remove organics, solids, and other ions to acceptable levels for release to the environment. Full details of the process model can be found in S1.

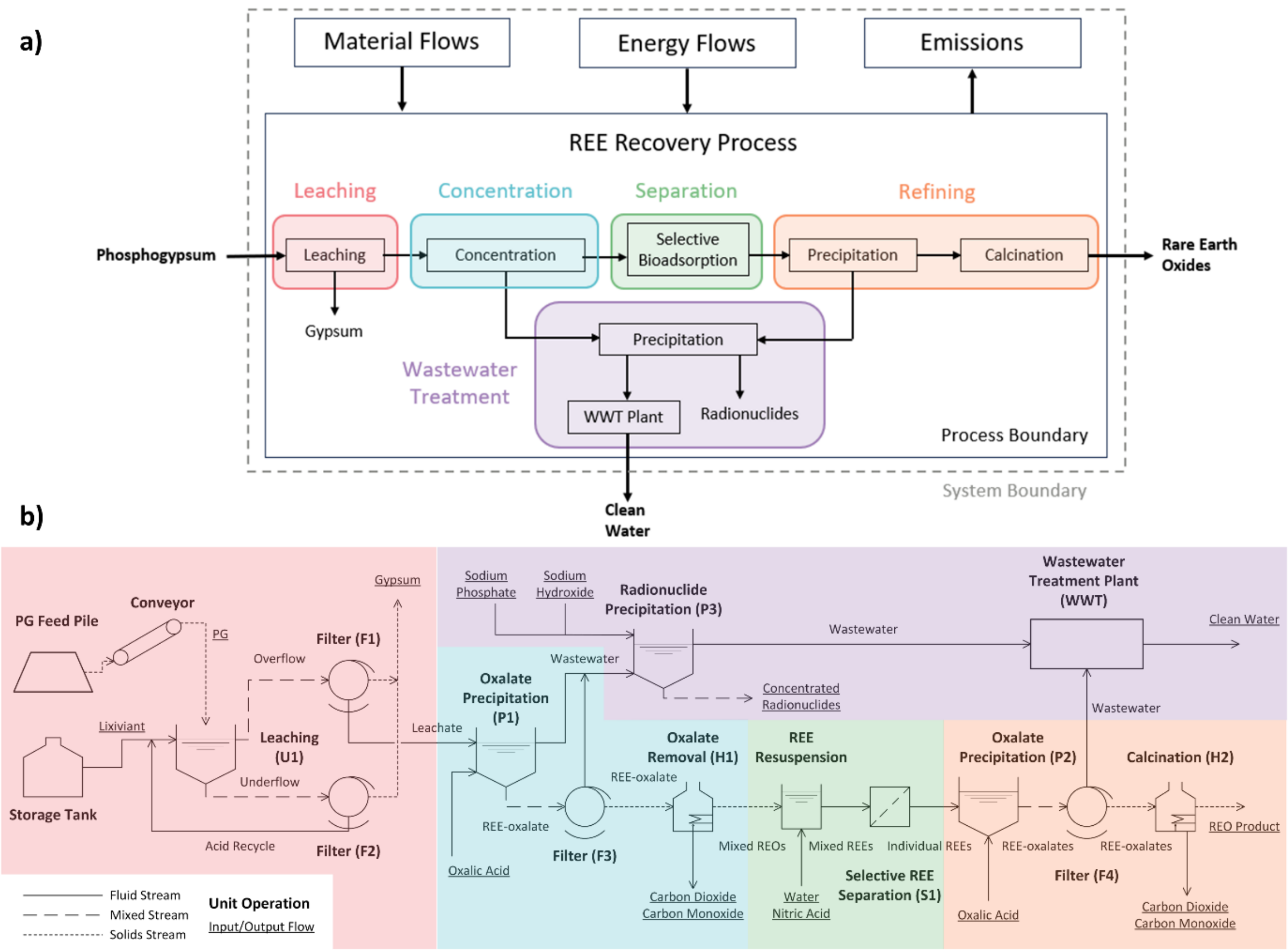

Figure 1: The **(a)** system diagram and **(b)** the process flow diagram for the rare earth recovery from phosphogypsum system (REEPS).

## 2.2. Life cycle assessment (LCA)

To quantify the environmental impact of the REEPS system, a cradle-to-gate LCA was performed following ISO 14040/14044 and current industry standards.[35,36] Two functional units were used: 1 kg of PG remediated (to assess the sustainability footprint of waste valorization compared to PG stacking) and 1 kg of REO produced (to assess the sustainability footprint of REE production compared to conventional REO production).

The goal of the LCA study is to compare the new REEPS system to conventional approaches, identify the hotspots, and combine with global sensitive analysis to determine the most influential process sections and parameters governing system-level sustainability. The system boundary (Figure 1a) includes impacts from raw material acquisition (cradle) through the production of the REO product (gate). System expansion approach was used for co-product handling. For the functional unit of PG remediated, the system was credited for the avoided production of gypsum and REOs. For the functional unit of REO produced, the system was credited for the avoided production of gypsum and the elimination of the PG waste. The system's direct emissions to water and air are quantified in addition to flows from raw material and chemical production. To model background processes, activities from the ecoinvent v3.9.1 cutoff database were used.[37] The US-SERC electricity grid was used to model the impact of electricity consumption due to the large volume of PG in this region. Impacts from construction and demolition were considered negligible and ignored. Transportation of PG to the system was not modeled in this study since there was no understanding of where the system should be built and what scale was optimal. In addition, the end of life for the concentrated radionuclide stream is unknown (e.g., storage, further refinement) and no impact was modeled for this flow.

To assess life cycle environmental impacts, the ReCiPe 2016 LCIA method (v1.03)[38] was chosen to characterize the REEPS system impacts. ReCiPe 2016 was chosen because it is widely used and includes a variety of impact categories relevant to this system. Specifically, ReCiPe 2016 comprehensively accounts for impacts from radioactive substances, toxicity, and land use which are relevant for mined rock leaching wastes (e.g., PG and acid mining tailings). Since PG and mining tailings are wastes stored for long or indefinite periods, the "long-term" version of each LCIA method was used.

## 2.3. Techno-economic analysis (TEA)

A TEA was performed to assess the profitability of the system by three indicators (net present value at a 15% interest rate (NPV15), internal rate of return (IRR), and minimum selling price (MSP)). The TEA uses a discounted cash flow rate of return analysis to assess profitability.[39] To calculate the capital investment, the purchase costs of equipment were estimated using cost correlations. Changes in equipment cost between scenarios were adjusted using calculated size factors.[39,40] Equipment costs were converted from past dollar values to 2022 U.S. dollars using the chemical engineering plant cost index. From these equipment costs, the total capital investment was calculated using the Lang factor method (Lang factor of 4.28). Capital was depreciated using the MACRS 7-year depreciation schedule. A plant life of 30 years with an uptime of 90%, a tax rate of 28%, and an interest rate of 15% was used for the analysis. Plant construction was completed in a 3-year period with 8%, 60%, and 32% of the capital expended in each year, respectively. Working capital was estimated as 5% of the fixed capital investment.[41]

Operating costs were calculated as the sum of variable and fixed operating costs. For variable operating costs, bulk chemical and utility prices were gathered from literature, government, and market sources (S3.1). The value of the REO product stream was estimated using a 'basket price' that considers the abundance of different REEs within PG (calculation in S3.2). The cost of the selective separation adsorbent was estimated as the sum of the price for specialized ion exchange resins and protein immobilized within this resin (calculation in S3.3). The cost of adsorbent replacements was annualized and included in the variable operating cost. The initial cost of the adsorbent was considered as the installed equipment cost for the selective separation (more discussion in S1.3.4). The producer price index was used to adjust prices to 2022 dollars. The fixed operating cost was calculated as the sum of labor, maintenance, and administrative costs.[39]

## 2.4. Uncertainty Characterization and Global Sensitivity Analysis

The foreground inventory was compiled by modeling the entire process (Figure 1b) in Python v3.10.13 (available at https://github.com/adsmer2/REEPS). This code leverages several established packages (e.g., brightway2[42], bioSTEAM[43,44], QSDsan[45]) to combine the process modelling, LCA, TEA, and Monte Carlo global uncertainty and sensitivity analysis into one tool. The Monte Carlo analysis used Latin Hypercube sampling (3000 samples) to evaluate the uncertainty and sensitivity. Other numbers of samples (500 and 1000) were compared to ensure that results were reproduceable (SI). The sensitivity of parameters was calculated using Spearman's rank correlation coefficients.[46] The Spearman rank coefficients were calculated separately for technological and contextual parameters because technological parameters can be controlled by engineers, while contextual parameters are dictated by external forces. By separating these two types of parameters, it becomes clearer which controllable parameters are most influential and should be the focus of future research. Parameters, assigned parameter uncertainty distributions, and the corresponding references for assigning these distributions are provided in S2 and S3.1.

# 3. Results and Discussion

## 3.1. Baseline Results

### 3.1.1. Pre-optimization of the system.

To ensure a more realistic estimate of profitability, the global optimal values of leaching parameters were identified prior to interpreting the results. The most profitable values of four key parameters were identified (bright yellow regions in Figure 2) for sulfuric acid leaching of PG (leaching temperature of 47°C, acid concentration 2.8 wt%, leaching time of 200 mins, and liquid-to-solid ratio (L/S ratio) of 275 wt% liquid). The black dots on Figure 2 represent the experimental local optimal conditions (experimental results provided in the Supporting Information, Section S4), which are different than the global, system optimal conditions determined here. The experimental data[23] was not extrapolated when finding the most profitable configuration. These optimal parameter values are only valid for this system since they represent the global optimal of the process. In addition, the limited data availability inhibited the consideration of interaction effects between parameters. Further details along with the complete set of the contour plots are available in S4.

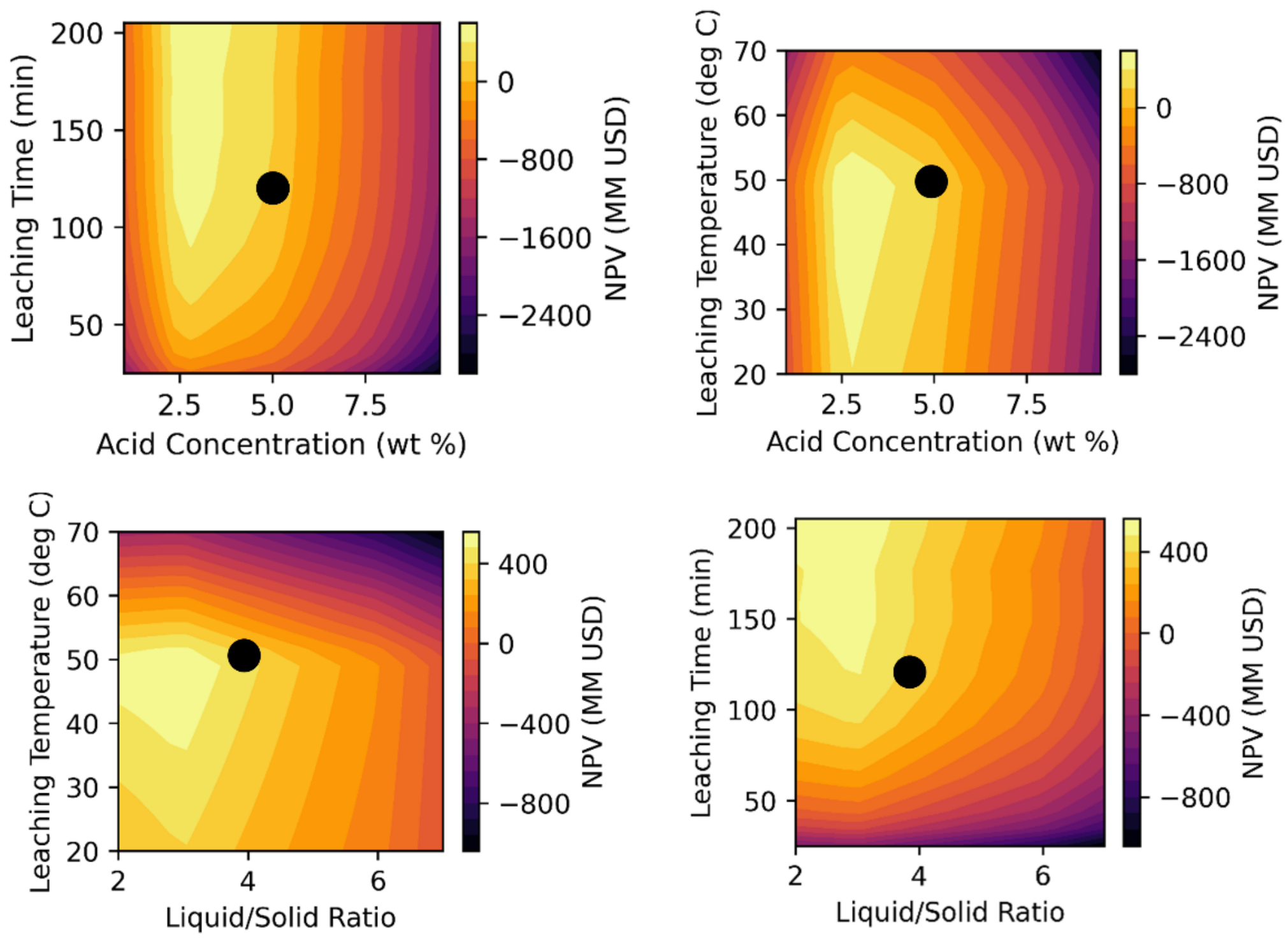

Figure 2: Contour plots showing how the values of key technological parameters were chosen to optimize net present value for the leaching unit. The black dot represents the experimentally determined local optimal conditions, which are different from the system's global optimal conditions (the bright yellow regions).

### 3.1.2. Baseline TEA Results

In the baseline scenario, the REEPS system has a return on investment (ROI) of 23% and a payback period of 4 years (for a 30-year investment). Considering the time value of money using a discounted cash flow analysis, the system is profitable (NPV15 above zero and an IRR above 15%). The high NPV ($570 million) suggests that the system can be profitable. However, the IRR of 20% is a result of high upfront costs from capital expenditure and reduced returns over the system lifetime. Further, the REEPS system has a high level of uncertainty due to the low technological readiness of the system, indicating that this system could be a risky investment compared to other investments. As observed in the probability density plot (Figure 5), the NPV15 ranges from around -$1 billion to $2 billion with the most probable value being around the baseline at $570 million. This uncertainty for the economic indicators is larger than what is observed for environmental indicators. This difference may be due to the uncertainty in environmental indicators being driven solely by technological parameters (e.g., leaching temperature), whereas economic indicators are subject to changes in both technological and contextual parameters (e.g., chemical prices, REO market prices, etc). Increasing the technological readiness of technologies relevant to REE recovery from PG is key to reducing the large uncertainties in system sustainability.

Though the REEPS system can be profitable, it is important to further understand what costs are driving the profitability to guide further research and process improvements. Figure 3**Error! Reference source not found.**b shows the breakdown of the costs contributing to the MSP ($44·kg$^{-1}$·REO) by process section. Approximately half of the MSP is from operating

costs ($23·kg$^{-1}$·REO, composed of chemicals and utilities) with the other half being related to capital recovery cost ($20·kg$^{-1}$·REO composed of capital depreciation, average income tax, and average return on investment). In addition, there is a smaller contribution due to fixed operating costs ($5.2·kg$^{-1}$·REO) and a credit from the sale of the byproduct gypsum (-$4.3·kg$^{-1}$·REO). The majority of the capital cost is from the ion exchange resin and the biomolecule ligand ($730 million and $240 million, respectively) in the selective separation ($15·kg$^{-1}$·REO). Replacements for the adsorbent comprise the majority of the chemical and material costs for the selective separation as well. The concentration and refining section costs ($2.9·kg$^{-1}$·REO and $1.2·kg$^{-1}$·REO, respectively) are primarily due to precipitant consumption, specifically oxalic acid. The leaching section costs ($4.1·kg$^{-1}$·REO) are split between capital, utilities, and chemicals ($8.3·kg$^{-1}$·REO) and the gypsum byproduct (-$4.3·kg$^{-1}$·REO). The chemical cost of leaching is primarily from sulfuric acid while the primary utility cost is for natural gas heating. The wastewater treatment section expenses ($12·kg$^{-1}$·REO) are split between operating costs ($9.6·kg$^{-1}$·REO) and capital costs ($2.0·kg$^{-1}$·REO). The primary operating cost is the sodium hydroxide for neutralizing the acid waste from the leaching section. Therefore, reducing acid use and selective separation costs should be a priority for improving the profitability of the system.

### 3.1.3. Baseline LCA results

When considering the functional unit of one kg of REO produced, the REEPS system has a mixed performance compared to conventional REO production (Figure 3a**Error! Reference source not found.**). We used both endpoint and midpoint impact assessment to understand system sustainability compared to conventional methods. Endpoint impact methods use value weighting to condense multiple midpoint impact categories (e.g., acidification and ecotoxicity) into one broad endpoint impact category (e.g., ecosystem quality). Therefore, endpoint analysis is helpful to reduce the complexity of analyzing tradeoffs but at the cost of detail and accuracy. The endpoint analysis shows that the REEPS system outperforms conventional REO production in ecosystem quality (93.0% of the impact) and resource depletion (96.2% of the impact) but underperforms in human health (213% of the impact). These advantages of the REEPS system are led by reductions in land use, eutrophication, ecotoxicity, human toxicity, and ionizing radiation. The advantages in ionizing radiation, human toxicity, and land use are largely due to the avoided impact from the stacking of PG waste (shown in Figure 4). Advantages in the other categories result from several high impact processes in conventional REO production that are not present in the REEPS system: storage of large volumes of acid mining tailings which can leach radioisotopes and toxic substances into the environment and toxic solvent use in the selective separation which contributes up to 30 % of the impact of conventional REO processing.[5]

Conversely, the REEPS system has higher impact (150-400%) on climate change, particulate matter formation, acidification, water use, and fossil depletion. Utilities account for 0.21% (ozone depletion) to 18% (energy resources) of the impact for the impact across selected categories. Overall, chemical consumption is the primary contributor to the environmental impacts of the system. For acidification, approximately 70% of the impact results from the consumption of sulfuric acid in leaching. Fossil depletion impact is largely due to the consumption of sodium hydroxide (73%) used for neutralization of acidic wastewater. Climate change, particulate matter formation, and acidification impacts are more evenly attributed to various chemicals used in the process (Figure 4). The REEPS photochemical oxidant formation and ozone depletion impacts are much higher than conventional REO production (approximately 940% and 1450% of the impact, respectively). These high impacts are almost exclusively due to oxalic acid consumption (approximately 92% and 83% of the impact, respectively). Therefore, future process modifications that reduce (or eliminate) the consumption of these chemicals, like oxalic acid, can greatly reduce the environmental impact of the REEPS system. Additionally, we compared impacts from REEPS without wastewater treatment (REEPS w/o WWT) to conventional REO production (cREO), since cREO systems mostly exclude wastewater treatment from the system scope. Notably, the climate change impact for REEPS w/o WWT and cREO were very similar (80 and 77 kg $CO_2$ eq.·kg$^{-1}$·REO, respectively). The full results are shown in Figure 3a.

When comparing to conventional PG stack treatment (functional unit of 1 kg of PG remediated), the REEPS system again has mixed performance. Endpoint analysis shows that the REEPS system has lower impacts than the PG stack treatment in ecosystem quality (97.5% of the impact) and resource depletion (-2.46 and 0 USD 2013·$kg^{-1}$·PG remediated, respectively), but higher impact on human health (1380% of impact). The main contributors to each impact category are the same as described above (the system operates identically regardless of which functional unit is considered). However, when comparing to the PG stack system, the avoided production of REOs from the conventional route is credited to the system. This credit is about the same magnitude as the impacts of the REEPS system for half the impact categories (Figure 4). For marine eutrophication, land use, and material resources, the credit is over 90% of the impact leading to large net negative impact for the REEPS system. However, for other categories (e.g., ozone depletion and photochemical oxidant formation), the credit is minimal (<10%) further emphasizing the need for process alternatives for the concentration section.

| Impact Category | REEPS [impact·kg$^{-1}$·REO] | REEPS (w/o WWT) [impact·kg$^{-1}$·REO] | Conventional REO (cREO)[a] [impact·kg$^{-1}$·REO] | Relative impact [% of cREO] | REEPS [impact·kg$^{-1}$·PG] | PG stack treatment (PGstack)[b] [impact·kg$^{-1}$·PG] | Relative impact [% of PGstack] |
|---|---|---|---|---|---|---|---|
| Acidification terrestrial | 1.32 | 1.19 | 0.407 | 325 | $2.00*10^{-3}$ | 0 | n.c. |
| Climate change | 115 | 80.0 | 76.5 | 150 | 0.110 | 0 | n.c. |
| Ecotoxicity freshwater | 10.4 | 8.53 | 12.6 | 82 | $4.63*10^{-3}$ | $1.05*10^{-3}$ | 440 |
| Ecotoxicity marine | 13.6 | 11.1 | 16.7 | 81 | $5.80*10^{-3}$ | $1.45*10^{-3}$ | 400 |
| Ecotoxicity terrestrial | 1110 | 950 | 1380 | 81 | 0.422 | $6.42*10^{-20}$ | n.c. |
| Energy resources | 27.8 | 19.5 | 18.0 | 155 | 0.028 | 0 | n.c. |
| Eutroph. freshwater | -0.080 | -0.101 | 0.036 | -225 | $3.61*10^{-5}$ | $2.4*10^{-4}$ | 15.0 |
| Eutroph. marine | 0.010 | $1.91*10^{-3}$ | 1.22 | 1.04 | $-1.43*10^{-3}$ | 0 | n.c. |
| Human Toxicity Carc. | 2.20 | -0.187 | 7.02 | 31 | $1.41*10^{-3}$ | $6.42*10^{-3}$ | 21.9 |
| Human Toxicity N-carc. | 205 | 162 | 333 | 62 | -0.010 | $3.41*10^{-2}$ | -29.6 |
| Ionising radiation | -286 | -289 | 13.2 | -2160 | $-4.21*10^{-3}$ | 0.568 | -0.742 |
| Land use | -41.9 | -42.6 | 27.2 | -154 | -0.030 | 0.085 | -35.0 |
| Material resources | 4.12 | -0.054 | 52.8 | 7.81 | -10.7 | 0 | n.c. |
| Ozone depletion | $6.32*10^{-4}$ | $6.00*10^{-4}$ | $6.35*10^{-5}$ | 995 | $1.14*10^{-6}$ | 0 | n.c. |
| Particulate matter | 0.410 | 0.335 | 0.152 | 267 | $5.64*10^{-4}$ | 0 | n.c. |
| Photochemical ox. human health | 1.65 | 1.56 | 0.197 | 837 | $2.87*10^{-3}$ | 0 | n.c. |
| Photochemical ox. ecosystems | 1.66 | 1.56 | 0.203 | 817 | $2.88*10^{-3}$ | 0 | n.c. |
| Water use | 1.75 | 0.909 | 1.39 | 127 | $1.66*10^{-3}$ | 0 | n.c. |
| Ecosystem quality [species*yr] | $4.36*10^{-7}$ | $2.62*10^{-7}$ | $4.68*10^{-7}$ | 93.0 | $8.94*10^{-10}$ | $9.17*10^{-10}$ | 97.5 |
| Human health [DALYs] | $4.19*10^{-4}$ | $3.23*10^{-4}$ | $1.97*10^{-4}$ | 213 | $4.66*10^{-7}$ | $3.39*10^{-8}$ | 1380 |
| Natural resources [USD 2013] | 8.59 | 5.91 | 8.93 | 96.2 | -2.46 | 0 | n.c. |

[a]Environmental impacts for conventional REO production were estimated using an activity from the ecoinvent 3.9.1 cutoff database.[37] The activity named “rare earth oxides production, from rare earth carbonate concentrate” for the CN-FJ region and with neodymium oxide as the reference product. To convert from a functional unit of 1 kg neodymium oxide to 1 kg of REO, impacts for each category were divided by the economic allocation factor (0.7323) for neodymium oxide.

[b]Environmental impacts for the conventional PG stack treatment were calculated using an adapted version of a published life cycle inventory.[24] Flows were updated to be compatible with the ecoinvent 3.9.1 cutoff database. Additional details are provided in S6.

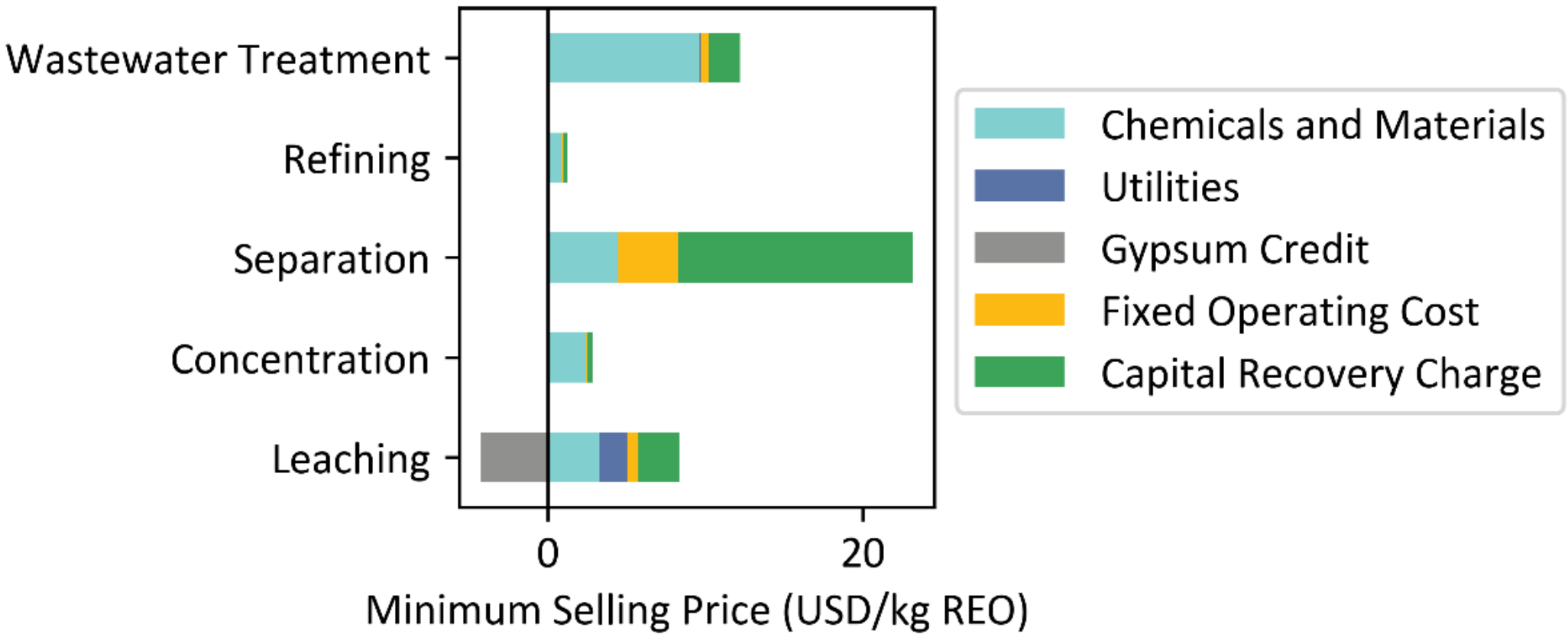

Figure 3: **(a)** Environmental impact of the REEPS system compared to conventional REO production and PG stack treatment using the ReCiPe 2016 LCIA method. The relative impact of the REEPS system compared to conventional techniques shows the extent of REEPS advantages (blue text) and disadvantages (red text). Midpoint category impacts are shown above the dotted line with endpoint impacts below the dotted line. **(b)** The contribution of each process section towards the minimum selling price of the REO product.

### 3.1.4. Hotspot analysis

Three process sections (leaching, concentration, and wastewater treatment) are responsible for over 75% of the environmental impacts (Figure 4). The primary impact from those process sections is from chemical consumption. The leaching section uses sulfuric acid in large volumes to extract the REEs from the solid gypsum matrix at elevated temperatures generated by burning natural gas. Though, a modest benefit is observed for the generation of a saleable gypsum coproduct (especially for particulate formation and land transformation). The concentration section requires a large excess of oxalic acid to bind metals in the leachate and to precipitate the REEs from the solution in preparation for the selective separation. The wastewater treatment section neutralizes the acidic wastewater from leaching and precipitates the rest of the toxic heavy metals with sodium hydroxide and sodium phosphate. Conventional REO production has a higher proportion of impacts from the selective separation since solvent extraction is inefficient requiring many stages and the use of high impact organic solvents and extractants. However, the selective separation in the REEPS system cannot fairly be compared since the model doesn't include the chemicals for regeneration or pumping costs (insufficient data). Though these impacts are expected to be minimal compared to solvent extraction.[47,48] The REEPS system also doesn't include further refining processes like electrolysis which contributes roughly 10-30% of the total impact of conventional REO production.[6] In this study it was assumed that the separation produced high enough purity REOs that further refining would not be necessary, but this will be important to revisit in future work using a more detailed separation model.

To avoid the impact of the most influential chemicals, changes to the REEPS process scheme should be explored. Specifically, improvements (or alternatives) to acid leaching for moving REEs from the solid phase to the liquid phase could reduce system impacts by up to 60%. Improvements in leaching could come from using acids with higher REE leaching efficiency and understanding, fundamentally, how acids and technical parameters influence this leaching efficiency. However, acid leaching not only reduces sustainability through chemical consumption and inefficient extraction, but also by requiring neutralization of the acidic wastewater now contaminated with toxic heavy metals. To mitigate this downstream problem, alternative 'leaching' techniques that do not require large volumes of acid and have high efficiency (e.g., ammono-carbonation)[22,29,49] are extremely promising. In addition, improved technologies for the concentration section could lead to an additional 20% reduction in impact. Due to the dilute nature of the feedstock, we included the concentration section to isolate and concentrate the REEs from the leachate to make the downstream separation more effective. However, it is currently unclear what specifications the concentration section must meet to enable an efficient selective separation. By developing detailed models for the selective separation, we can clarify the requirements of the concentration section and further optimize the system. Depending on the required specifications, we can also consider alternative concentration technologies that reduce chemical consumption (e.g., electrodialysis, filtration) to identify more sustainable process schemes. Similarly, we can examine other selective separations (e.g., membrane adsorption, solvent extraction, nanofiltration). By expanding the established Python framework (https://github.com/adsmer2/REEPS), we can 'plug and play' these different technologies to find more optimal process schemes and to examine trade-offs between technologies.

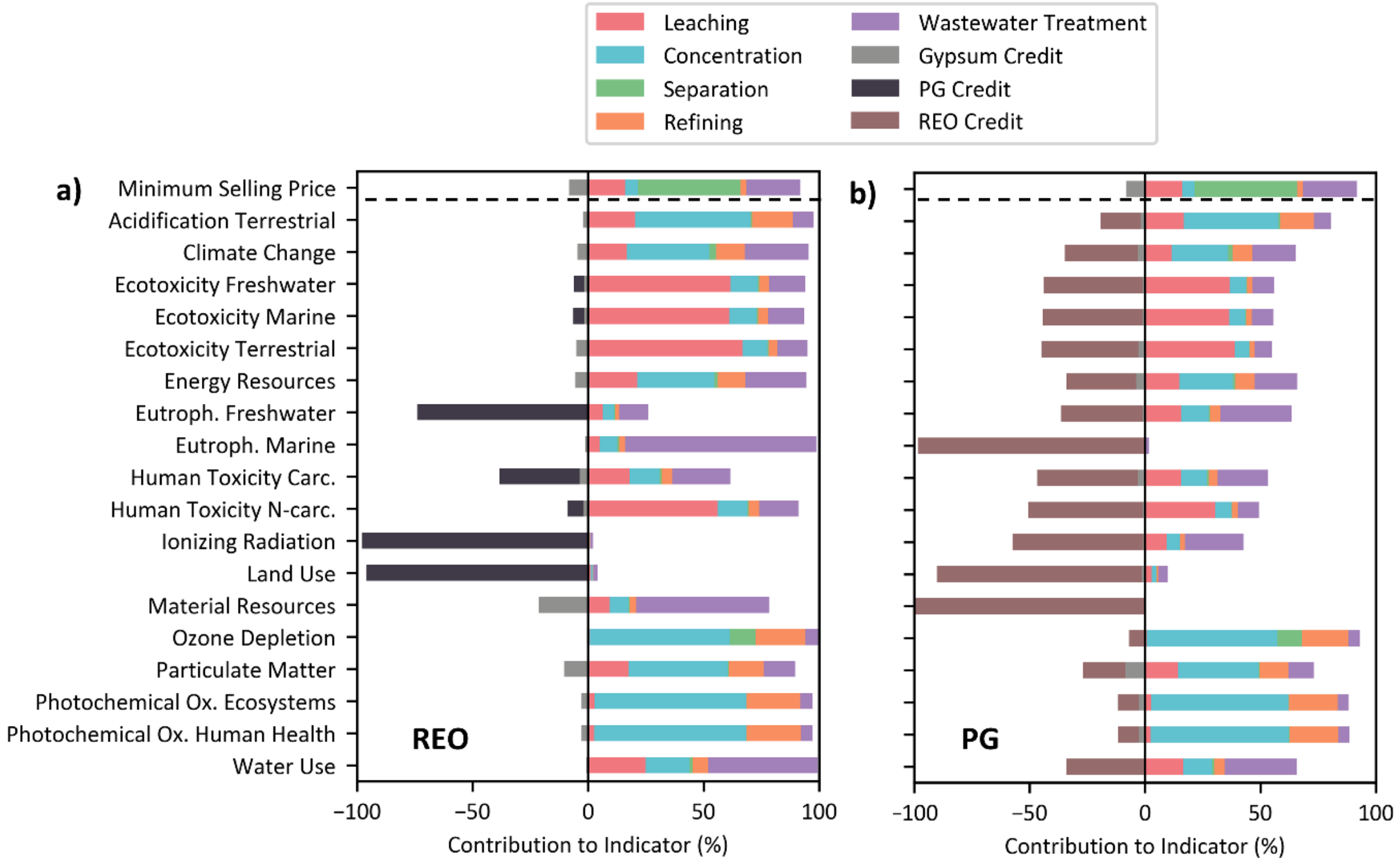

Figure 4: The contribution of each process section towards the environmental and economic impacts of the system using a functional unit of **(a)** 1 kg of REO produced and **(b)** 1 kg of PG remediated.

## 3.2. Probabilistic Sustainability

Sustainability assessments inherently involve uncertainties, stemming from variability in raw material inputs, process efficiencies, market fluctuations, and evolving regulatory frameworks. Traditional life cycle assessment (LCA) and techno-economic analysis (TEA) often rely on deterministic modeling, where fixed parameter values lead to point estimates that may not fully capture variability, particularly for early-stage technologies. Since the aim of this work is to guide decision-making across different REE recovery technologies, we focus on the foreground uncertainties across the technological space. This will help stakeholders compare technologies and set appropriate R&D targets.

Here, we introduce the concept of "Probabilistic Sustainability" for assessing and evaluating the potential of early stage/emerging technologies – especially for low technological readiness levels (TRL).[33,50] A Probabilistic Sustainability Assessment (PSA) framework is proposed, we incorporate system pre-optimization, uncertainty quantification and stochastic modeling to enhance decision-making robustness, making sustainability assessments more reflective of uncertainties in future real-world implementation. More importantly, this approach avoids false precision in decision-making and provides critical insights for policymakers to support more informed and resilient sustainability strategies.

### 3.2.1 Uncertainty Analysis

Since no large-scale system exists for REE recovery from PG, we first needed to identify the possible design space of the system. We defined the feedstock PG REE content, 0.5 wt%, by taking the median REE content of PG stacks globally (ranging from 0.02-0.9 wt% REE).[22] Next, we used this REE content to calculate a reasonable plant capacity (approximately 1 M kg PG processed·hr$^{-1}$). We ensured that the annual REO production rate was less than the global demand for REOs. Further, we confirmed that a PG containing region has a PG supply that could meet this REO production rate (S1.2). Additionally, we considered that some fraction of the REEs within PG would be unrecoverable using the current process scheme (due to the highly dilute nature of the feed source). Therefore, REEs with relative abundance below 1 wt% (Sm, Tb, Eu, Ho, Yb, Lu, Y, and Sc) were not modeled as part of the REE stream for the analysis. This consideration led to a loss in REE product of 2.2 wt% and reduced the value of the combined REO product from $55.0·kg$^{-1}$·REO to $51.5·kg$^{-1}$·REO. Using the above REE content, plant capacity, and recoverability, we defined a baseline scenario for REE recovery from PG. The baseline scenario results for 10 key sustainability indicators are shown in Figure 5. A figure including all indicators for each functional unit is provided in S5.

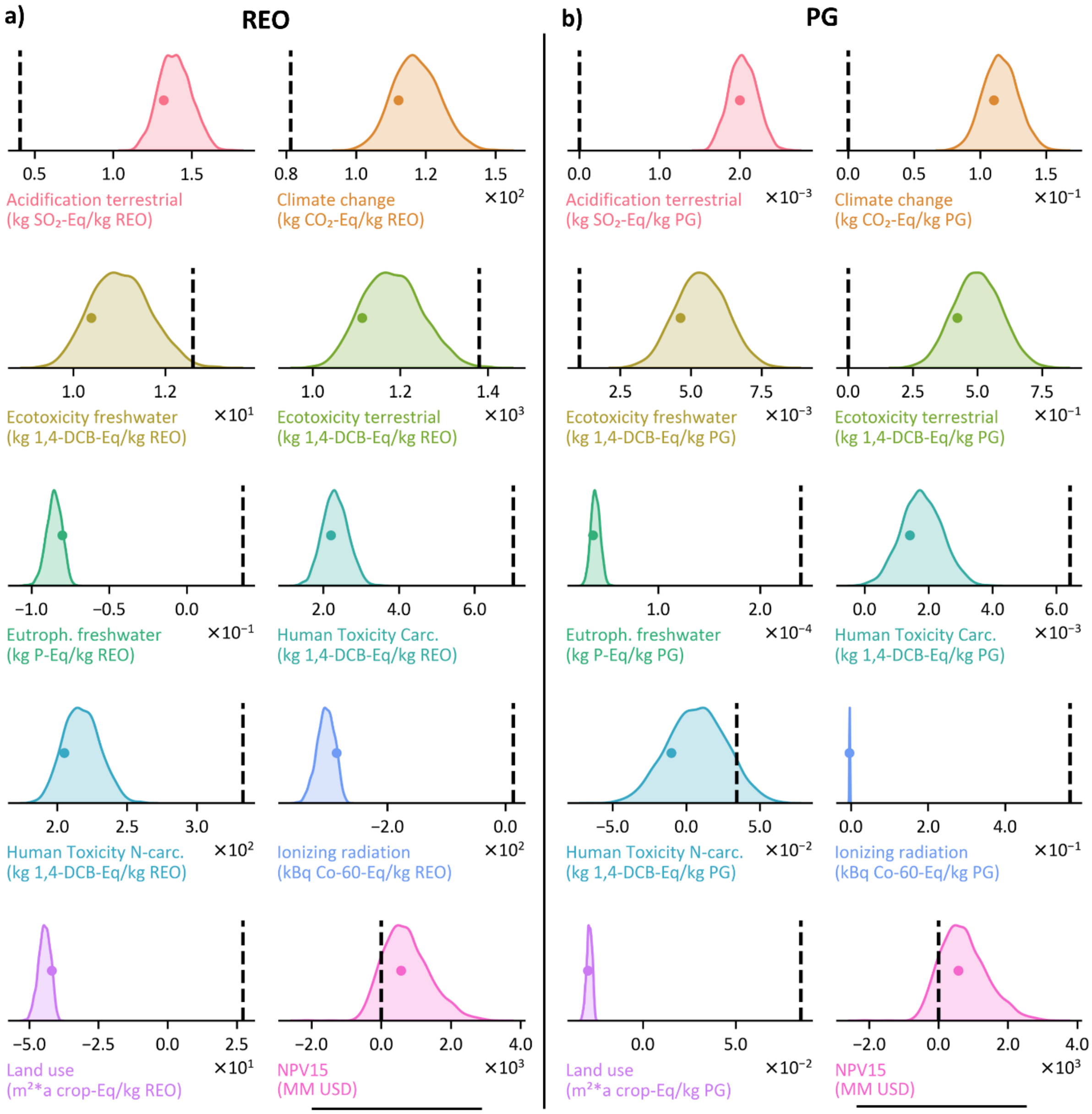

Figure 5: The baseline value (**dots**) and the probability density (**shaded regions**) of indicator values for the two functional units: **(a)** 1 kg of REO produced and **(b)** 1 kg of PG remediated. Peaks represent a higher probability of an indicator value, while broad distributions represent greater uncertainty. The indicator values of conventional REO production (**dotted black line**) are shown for comparison (with NPV15 > 0 indicating profitability). The indicator underlined with a **solid black bar** is an economic indicator (net present value at an interest rate of 15%) evaluated by TEA. All other indicators were evaluated by LCA using the ReCiPe 2016 LCIA method. The probability densities were calculated by collecting 3000 samples of the system using Latin Hypercube sampling from defined parameter distributions (S2 and S3.1). The full figure with all environmental and economic indicators can be found in S5.

### 3.2.2. Global Sensitivity Analysis

We used a global sensitivity analysis to identify key parameters for process improvement (Figure 6). The list of parameters along with their uncertainty distributions are provided in S2 and S3.1. A figure showing the sensitivity for the other functional unit is in S4. Of the technological parameters (Figure 6a), seven parameters had coefficients larger than 0.15: REE recovery of the selective separation (S1), sodium hydroxide required for neutralizing the wastewater (P3), oxalic acid for precipitation of REEs (P1 and P2), the concentration of acid for leaching (U1), the leaching overflow to underflow ratio (U1), and the leaching lixiviant to solid ratio (U1). The process unit operation codes are given in Figure 1b. One of the main drivers for the sensitivity of these parameters is their influence on the production of REOs. The REO product is very valuable and has a large environmental impact (avoided production credit) leading to increased sustainability as more REOs are produced. Therefore, minimizing waste from the selective separation (i.e., REE recovery) should prioritized when designing the unit. Similarly, leaching efficiency should be prioritized in the leaching unit. Developing a greater understanding of what acids and parameters (e.g., acid concentration and solvent-to-solid ratio) are most effective should be prioritized. Further, we need a better understanding of how scaling up bench-scale leaching experiments influences leaching efficiency (batch leaching at lab scale is unlikely to perform as well as large countercurrent flow systems that optimize thermodynamic driving forces). Understanding how operational parameters in these larger flow systems (e.g., overflow to underflow ratio) will also be important in reducing uncertainty in the process design. Another main driver for the sensitivity of the seven parameters is raw material consumption. Four of the parameters directly relate to material consumption: oxalic acid consumption in P1 and P2, sodium hydroxide consumption in P3, and acid concentration in U1. Therefore, using less of these chemicals, different chemicals, or different technologies altogether would lead to significant decreases in environmental impact and cost.

Regarding contextual parameters (Figure 6b), only economic indicators are relevant since changes in prices and TEA parameters do not affect environmental indicators. Interest rate and REO price are used to calculate IRR and MSP, respectively. Therefore, there is no sensitivity of these indicators to their respective parameters. Five contextual parameters have coefficients above 0.15: income tax rate, interest rate, sodium hydroxide price, REO price, biomolecule price, and number of operating days per year. Of these parameters, the interest rate and REO price are most influential to profitability. The interest rate is influential due to the high capital expenditure in early years making this investment riskier. As observed in Figure 4, reduction in the cost of the adsorbent should be prioritized to reduce sensitivity to this parameter by reducing capital cost. The REO price is also influential due to the wide range given to the distribution (36.1-67.0 $\$\cdot kg^{-1}\cdot$REO). Within the past decade REO prices have been highly volatile. Since the current REO supply chain still exists in a similar form, we made the distribution reflect this volatility.[2] In addition to this volatility in REO prices, we also considered that increased demand for the green energy transition may lead to higher REO prices in the future (reflected by the positive correlation to profitability in Figure 6b). However, we acknowledge the possibility of REO prices decreasing. Since REOs are produced together but have different market demand, it means that REOs with the least demand will be produced in excess (decreasing the price).[17] Similar to the large uncertainty in REO prices, the price of producing bulk biomolecules is highly uncertain. Some studies report theoretical bulk prices of proteins and peptides. However, more work is required to understand how much prices could vary for different molecules. The uncertainty range in this study ranges from the theoretical cost of bulk protein ($\$0.004\cdot g^{-1}$) to bulk peptide ($\$10\cdot g^{-1}$) with a more average number ($\$0.5\cdot g^{-1}$) chosen for the base case value.

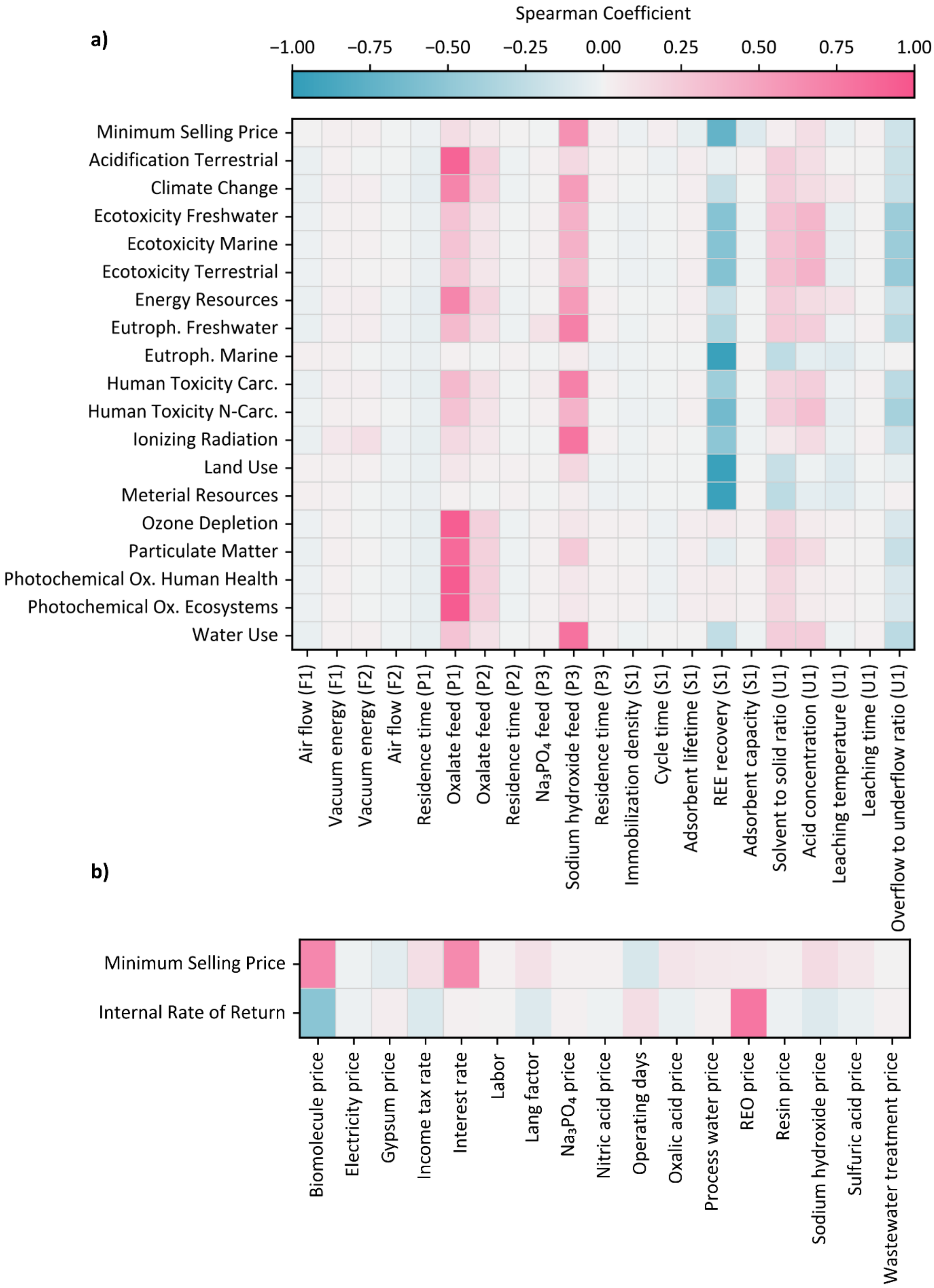

Figure 6: the sensitivity of environmental and economic indicators to **(a)** technological parameters and **(b)** contextual parameters. Blue indicates better performance (reduced environmental impact and higher profitability) with an increase in parameter value, while red indicates inferior performance as the parameter increases. The more vibrant the color, the greater the sensitivity of an indicator (y-axis) to a change in a parameter (x-axis). The indicator above the dotted black line is an economic indicator (evaluated by TEA) and indicators below are environmental indicators (evaluated by LCA for the functional unit of 1 kg of PG remediated). Sensitivity was calculated using Spearman rank correlations with Latin Hypercube sampling (3000 samples) of parameter distributions (given in S2 and S3.1).

### 3.2.3. Scenario analysis

We used a scenario analysis to explore which combination of plant capacity and PG REE content enable a profitable operation (Figure 7). In general, profitability increases (shown by decreasing MSP) as capacity increases. This increase in profitability with capacity indicates that larger centralized facilities should be prioritized over decentralized modular systems. However, a centralized facility would require transportation of PG waste to the processing plant. The implications of transportation on the sustainability of the system are not considered in this work. Future work will need to consider the logistics, cost, and environmental impact of transporting PG and identify regions that are ideal for a REE recovery facility. Regions that have large volumes of PG in close proximity (e.g., Florida with >1 billion tons in 24 stacks)[20] are promising for investigation.

However, the current REEPS system is only profitable for PG stacks with an REE content above roughly 0.5 wt%. To determine conditions that make operations with more dilute feedstocks profitable, we considered a few scenarios. First, we examined the effect of increased sales. If the value of REOs increases in the future (e.g., REO prices double), PG stacks with as little as 0.2 wt% REE content become profitable (Figure 7a). Another form of additional income could come from the producers or managers of PG waste. They may prefer to pay for remediation as opposed to having to manage a PG stack indefinitely with its risk of release. Second, we considered a reduction in capital cost. Even with improvements in the capacity of the separation unit (0.00577 to 0.025 mol·$L^{-1}$ adsorbent), the most dilute PG stacks remain inaccessible (Figure 7b). These results indicate that a new process scheme must be developed to be able to access the most dilute PG stacks (0.02-0.1 wt%). New process schemes should reduce system costs by utilizing new technologies and address the most sensitive parameters. Specifically, the cost of acid for REE leaching and subsequent neutralization of the wastewater is the primary barrier to recovery from dilute sources. A technology that disrupts this paradigm would greatly increase the viability of REE recovery from dilute minerals.

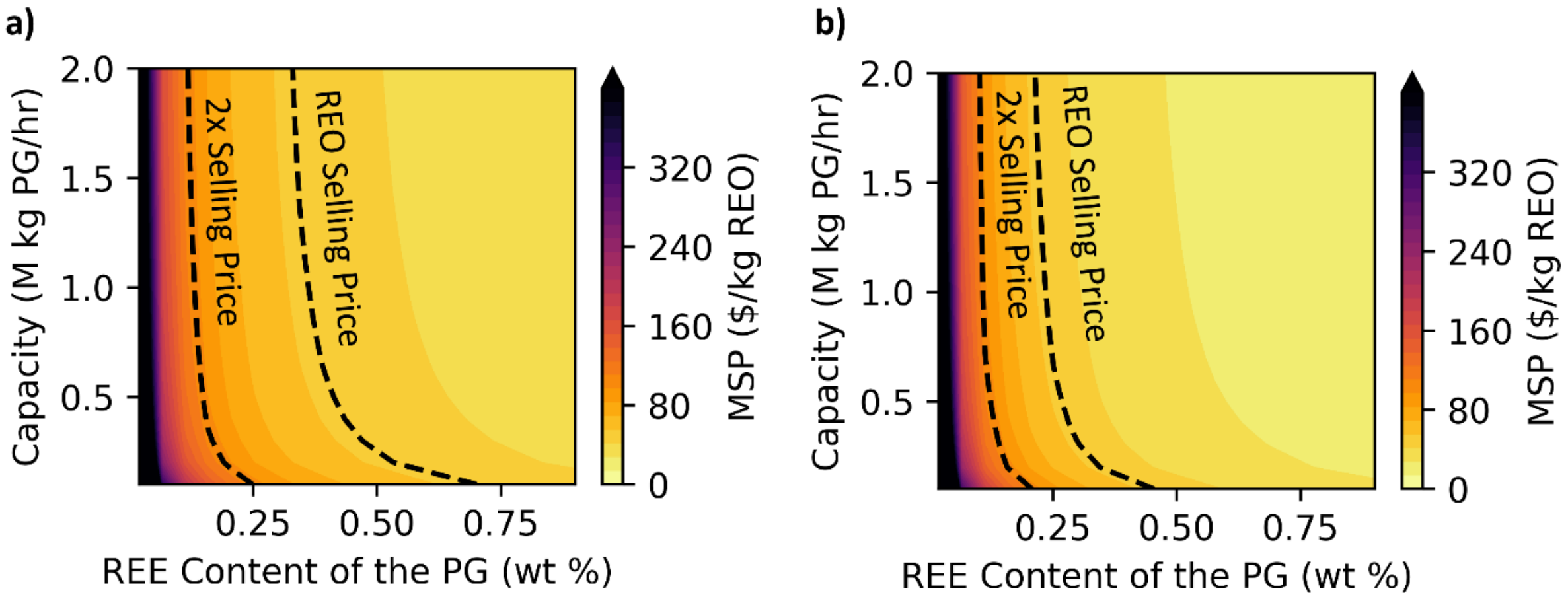

Figure 7: The effect of capacity and REE content in the PG on process profitability (measured by MSP) **(a)** for the baseline system and **(b)** considering improvements in the separation technology through increased capacity (from 0.0058 to 0.025 mol REE·$L^{-1}$ resin) and REE recovery (from 99 to 100%). Brighter regions indicate more profitable scenarios (lower MSP) where all contours to the right of the dotted lines are profitable. The dotted lines are the current basket REO selling price ($51.5·$kg^{-1}$·REO) and a future scenario with two times the current selling price ($103·$kg^{-1}$·REO). The REE content of PG stacks around the globe varies between 0.02-0.9 wt% REE.[22]

# 4. Limitations and path forward

There are several limitations to the results of this study, which are consistent with those typically observed in critical mineral recovery sustainability analysis work and are inherent to the current methodologies commonly used in this field.

First, more reliable, relevant, and complete experimental data would increase certainty in the REEPS system results (e.g., leaching data for more acids and multiple PG sources). Considering process alternatives for these technologies must also be explored. Second, a better understanding of the life cycle impacts of conventional PG stack treatment and REO production is required. Many of the technical details of conventional REE production are not public information.[6,7] Second, the illegal (and therefore unregulated) production of REEs, often occurring across various countries outside of US, makes it difficult to assess the full environmental impact of the current REO supply chain.[51] A qualitative discussion of the limitations and uncertainty of LCAs for REO production is available in the literature, but uncertainty has not been rigorously quantified to date.[7,52]

Further, studies of REE systems report results in terms of either a mixed REE product, a purified individual REO product (either as an individual REO like Nd or in total), or a refined rare earth metal product. These three different purity REE products, with greatly different market values, are a result of different system scopes making comparison between studies challenging. Due to uncertainty in results from other conventional REO studies (e.g., functional unit, allocation, scope), we compared environmental impacts to ecoinvent. However, even the ecoinvent results are not entirely transparent with their allocation. Ecoinvent performs mass-based subdivision of the system prior to economic allocation, but they do not describe how they choose which REOs are reference products (for subdivision) and which are allocateable byproducts (for economic allocation). Therefore, we converted the ecoinvent results from a functional unit of individual separated REOs (e.g., $Nd_2O_3$) to total separated REOs for comparison to the total amount of separated REOs considered here. We do acknowledge that a mixture of separated REOs has different value due to its composition. However, we recommend using this total separated REO functional unit for comparison to avoid uncertainties due to allocation, especially for feedstocks of similar composition like we have here.

For PG stack treatment, only two studies have quantified the environmental impact.[24,25] Of these, only one provided a detailed life cycle inventory.[24] These two studies disagreed on the primary impact of PG (e.g., respiratory inorganics is the highest impact in one study and has no impact in the other) suggesting we do not understand what flows to consider and how different geographies affect LCA results. The LCA methodology is not effective at quantifying the long-term impacts of stored wastes on the environment. Decisions on how to treat long-term impacts from mining have been shown to change impact by up to eight orders of magnitude.[53] This limitation must be resolved moving forward as it restricts our ability to consider temporally relevant impacts to accurately assess the sustainability of technologies.[53,54]

Future efforts in LCIA methodologies in this field should also improve modeling methods that quantify the impacts of radioactivity and water depletion. For radioactivity, current LCIA methods calculate the direct impact of radioactive substances on human health but do not consider the indirect impact of these radionuclides on ecosystems. In addition, current LCIA methods don’t have a way to reliably link impacts to how radionuclide concentrations change as they move through the environment. Some models outside of the LCA field exist that could be integrated into LCIA methodologies to more accurately quantify the impact of radionuclide releases on organism and ecosystem health.[55] For impacts of water depletion, most LCIA methods consider only water flows that do not return to the native aquifer.[56] However, water reclamation systems, like REEPS, are interested in returning water back to the native aquifer as part of their function. Therefore, it is important to note that the water use impact used here, along with those reported in other current REE recovery studies, is not comprehensive. The REEPS system could theoretically have a net negative ‘water depletion’ impact since water is being returned to the environment. Addressing these limitations should be a priority in the future development of LCA methodologies for the sustainability assessments of REE recovery technologies.

# 5. Conclusion

We established open-access modeling tools to translate REE separation concepts to field-scale systems and evaluated the potential economic and environmental feasibility under uncertainty. This work will also enable the community to standardize sustainability assessments of REE recovery and purification concepts and prioritize R&D pathways.

Recovery of rare earth elements from phosphogypsum can be profitable (IRR of >15% in 87% of baseline scenario simulations) and reduce the environmental impact compared to conventional methods of REO production and PG treatment. An integrated LCA and TEA was performed with global uncertainty and sensitivity analysis to assess the sustainability of the system for two functional units: 1 kg of REO and 1 kg of PG remediated. Compared to conventional REO production, the REEPS system had lower impacts to ecosystem quality (93.0% of the impact) and resource depletion (96.2%), but a higher human health impact (213%). Of the midpoint categories, 8 were conclusively worse, 9 conclusively better, and one within uncertainty of conventional REO production. Compared to conventional PG stack treatment, the REEPS system again showed reductions in impacts for ecosystem quality (97.5% of the impact) and resource depletion (0 and -2.46 USD 2013·$kg^{-1}$·PG remediated, respectively), but higher impact on human health (1380% of the impact). Of the midpoint categories, 11 were conclusively worse, 8 conclusively better, and one within uncertainty of conventional PG stack treatment. These uncertainties do not account for uncertainty in conventional process impacts nor the implementation of technology alternatives in the system but do provide insight into potential system improvements.

In the REEPS system, the primary contributor to these environmental impacts is chemical consumption (>82%), with sulfuric acid, sodium hydroxide, and oxalic acid having the most significant impact. In addition, the cost of the REEPS system ($44·$kg^{-1}$·REO) was broken down by process section. The largest cost contributors were the chemical consumption in the leaching and wastewater treatment sections ($3.3·$kg^{-1}$·REO and $10·$kg^{-1}$·REO, respectively) and the cost of the bio-based adsorbent used in the selective separation ($23·$kg^{-1}$·REO). The coproduction of gypsum from sulfuric acid leaching showed a modest improvement in profitability ($4.3·$kg^{-1}$·REO) and environmental impact (up to 25% for some impact categories) indicating that the production of coproducts is an important avenue for increasing sustainability. A scenario analysis showed that the current process scheme is profitable for capacities above 100,000 kg PG·$hr^{-1}$ and PG REE contents above 0.5 wt%. However, even with improvements to the selective separation technology, the most dilute sources of PG (0.02-0.1 wt% REE) remain inaccessible necessitating improvements in the process scheme and performance of other process technologies.

Future work must examine ways to increase REE recovery, reduce chemical consumption, and reduce the cost of selective REE separations. In the bio-adsorbent separation used here, half the cost comes from the agarose resin itself and could be reduced by attaching ligands to less expensive substrates. One challenge of the current protein-resin coupling scheme is the requirement of an aminated agarose resin for one specific click-chemistry reaction to couple the protein to the solid. These specialty resins have high costs of approximately $45·$L^{-1}$. Pursuing alternative coupling reactions between the proteins and inexpensive substrates (resins or fiber mats) could have a meaningful impact on the economics of the adsorption process. In the leaching and wastewater treatment sections, acid and base consumption have high environmental impact (approximately 60% of the impact across multiple impact categories) and limit the profitability to PG stacks with REE contents above 0.5 wt%. Therefore, exploring the use of different leaching acids (e.g., nitric acid, hydrochloric acid), new lixiviants that increase REE extraction, or alternative leaching technologies (e.g., ammono-carbonation) is vital to creating a more sustainable system. In summary, this work evaluated the sustainability of a potential route for REE recovery from the secondary source, PG, and identified directions for technological and process improvements. These improvements are critical for establishing a robust REO supply chain using dilute secondary sources, which is becoming increasingly important as the demand for REOs increases during the clean energy transition.

## Conflict of Interest

The authors report no conflict of interest.

## Acknowledgements

We thank the National Science Foundation for their support under award number ECO-CBET-2133530.

# References

(1) IPCC; Masson-Delmotte, V.; Zhai, P.; Pörtner, H.-O.; Roberts, D.; Skea, J.; Shukla, P.; Pirani, A.; Moufouma-Okia, W.; Péan, C.; Pidcock, R.; Connors, S.; Matthews, R.; Chen, Y.; Zhou, X.; Gomis, M.; Lonnoy, E.; Maycock, T.; Tignor, M.; Tabatabaei, M. *Global Warming of 1.5°C. An IPCC Special Report on the Impacts of Global Warming of 1.5°C above Pre-Industrial Levels and Related Global Greenhouse Gas Emission Pathways, in the Context of Strengthening the Global Response to the Threat of Climate Change, Sustainable Development, and Efforts to Eradicate Poverty*; 2018. https://doi.org/10.1017/9781009157940.

(2) *The Role of Critical Minerals in Clean Energy Transitions*; IEA: Paris, 2021. https://www.iea.org/reports/the-role-of-critical-minerals-in-clean-energy-transitions.

(3) Gupta, C. K.; Krishnamurthy, N. Extractive Metallurgy of Rare Earths. *International Materials Reviews* **1992**, *37* (1), 197–248. https://doi.org/10.1179/imr.1992.37.1.197.

(4) Shahbaz, A. A Systematic Review on Leaching of Rare Earth Metals from Primary and Secondary Sources. *Minerals Engineering* **2022**, *184*, 107632. https://doi.org/10.1016/j.mineng.2022.107632.

(5) Vahidi, E.; Zhao, F. Environmental Life Cycle Assessment on the Separation of Rare Earth Oxides through Solvent Extraction. *Journal of Environmental Management* **2017**, *203*, 255–263. https://doi.org/10.1016/j.jenvman.2017.07.076.

(6) Zapp, P.; Schreiber, A.; Marx, J.; Kuckshinrichs, W. Environmental Impacts of Rare Earth Production. *MRS Bulletin* **2022**, *47* (3), 267–275. https://doi.org/10.1557/s43577-022-00286-6.

(7) Schreiber, A.; Marx, J.; Zapp, P. Life Cycle Assessment Studies of Rare Earths Production - Findings from a Systematic Review. *Science of The Total Environment* **2021**, *791*, 148257. https://doi.org/10.1016/j.scitotenv.2021.148257.

(8) Opare, E. O.; Struhs, E.; Mirkouei, A. A Comparative State-of-Technology Review and Future Directions for Rare Earth Element Separation. *Renewable and Sustainable Energy Reviews* **2021**, *143*, 110917. https://doi.org/10.1016/j.rser.2021.110917.

(9) El Ouardi, Y.; Virolainen, S.; Massima Mouele, E. S.; Laatikainen, M.; Repo, E.; Laatikainen, K. The Recent Progress of Ion Exchange for the Separation of Rare Earths from Secondary Resources – A Review. *Hydrometallurgy* **2023**, *218*, 106047. https://doi.org/10.1016/j.hydromet.2023.106047.

(10) Dong, Z.; Mattocks, J. A.; Deblonde, G. J.-P.; Hu, D.; Jiao, Y.; Cotruvo, J. A. Jr.; Park, D. M. Bridging Hydrometallurgy and Biochemistry: A Protein-Based Process for Recovery and Separation of Rare Earth Elements. *ACS Cent. Sci.* **2021**, *7* (11), 1798–1808. https://doi.org/10.1021/acscentsci.1c00724.

(11) Dong, Z.; Mattocks, J. A.; Seidel, J. A.; Cotruvo, J. A.; Park, D. M. Protein-Based Approach for High-Purity Sc, Y, and Grouped Lanthanide Separation. *Separation and Purification Technology* **2024**, *333*, 125919. https://doi.org/10.1016/j.seppur.2023.125919.

(12) Li, Z.; Diaz, L. A.; Yang, Z.; Jin, H.; Lister, T. E.; Vahidi, E.; Zhao, F. Comparative Life Cycle Analysis for Value Recovery of Precious Metals and Rare Earth Elements from Electronic Waste. *Resources, Conservation and Recycling* **2019**, *149*, 20–30. https://doi.org/10.1016/j.resconrec.2019.05.025.

(13) Azimi, G.; Sauber, M. E.; Zhang, J. Technoeconomic Analysis of Supercritical Fluid Extraction Process for Recycling Rare Earth Elements from Neodymium Iron Boron Magnets and Fluorescent Lamp Phosphors. *Journal of Cleaner Production* **2023**, *422*, 138526. https://doi.org/10.1016/j.jclepro.2023.138526.

(14) Chowdhury, N. A.; Deng, S.; Jin, H.; Prodius, D.; Sutherland, J. W.; Nlebedim, I. C. Sustainable Recycling of Rare-Earth Elements from NdFeB Magnet Swarf: Techno-Economic and Environmental Perspectives. *ACS Sustainable Chem. Eng.* **2021**, *9* (47), 15915–15924. https://doi.org/10.1021/acssuschemeng.1c05965.

(15) Alipanah, M.; Park, D. M.; Middleton, A.; Dong, Z.; Hsu-Kim, H.; Jiao, Y.; Jin, H. Techno-Economic and Life Cycle Assessments for Sustainable Rare Earth Recovery from Coal Byproducts Using Biosorption. *ACS Sustainable Chem. Eng.* **2020**, *8* (49), 17914–17922. https://doi.org/10.1021/acssuschemeng.0c04415.

(16) Fritz, A. G.; Tarka, T. J.; Mauter, M. S. Technoeconomic Assessment of a Sequential Step-Leaching Process for Rare Earth Element Extraction from Acid Mine Drainage Precipitates. *ACS Sustainable Chem. Eng.* **2021**, *9* (28), 9308–9316. https://doi.org/10.1021/acssuschemeng.1c02069.

(17) Fritz, A. G.; Tarka, T. J.; Mauter, M. S. Assessing the Economic Viability of Unconventional Rare Earth Element Feedstocks. *Nat Sustain* **2023**, *6* (9), 1103–1112. https://doi.org/10.1038/s41893-023-01145-1.

(18) Jyothi, R. K.; Thenepalli, T.; Ahn, J. W.; Parhi, P. K.; Chung, K. W.; Lee, J.-Y. Review of Rare Earth Elements Recovery from Secondary Resources for Clean Energy Technologies: Grand Opportunities to Create Wealth from Waste. *Journal of Cleaner Production* **2020**, *267*, 122048. https://doi.org/10.1016/j.jclepro.2020.122048.
(19) US EPA, O. *TENORM: Fertilizer and Fertilizer Production Wastes*. https://www.epa.gov/radiation/tenorm-fertilizer-and-fertilizer-production-wastes (accessed 2022-10-26).
(20) *Phosphogypsum Stacks*. https://fipr.floridapoly.edu/about-us/phosphate-primer/phosphogypsum-stacks.php (accessed 2023-12-21).
(21) *U.S.: rare earths consumption 2022*. Statista. https://www.statista.com/statistics/616726/consumption-of-rare-earths-in-the-united-states/ (accessed 2024-01-11).
(22) Mukaba, J.-L.; Eze, C. P.; Pereao, O.; Petrik, L. F. Rare Earths' Recovery from Phosphogypsum: An Overview on Direct and Indirect Leaching Techniques. *Minerals* **2021**, *11* (10), 1051. https://doi.org/10.3390/min11101051.
(23) Liang, H.; Zhang, P.; Jin, Z.; DePaoli, D. Rare Earths Recovery and Gypsum Upgrade from Florida Phosphogypsum. *Mining, Metallurgy & Exploration* **2017**, *34* (4), 201–206. https://doi.org/10.19150/mmp.7860.
(24) Tsioka, M.; Voudrias, E. A. Comparison of Alternative Management Methods for Phosphogypsum Waste Using Life Cycle Analysis. *Journal of Cleaner Production* **2020**, *266*, 121386. https://doi.org/10.1016/j.jclepro.2020.121386.
(25) Kulczycka, J.; Kowalski, Z.; Smol, M.; Wirth, H. Evaluation of the Recovery of Rare Earth Elements (REE) from Phosphogypsum Waste – Case Study of the WIZÓW Chemical Plant (Poland). *Journal of Cleaner Production* **2016**, *113*, 345–354. https://doi.org/10.1016/j.jclepro.2015.11.039.
(26) Rychkov, V.; Kirillov, E.; Kirillov, S.; Bunkov, G.; Botalov, M.; Semenishchev, V.; Smyshlyaev, D.; Malyshev, A.; Taukin, A.; Akcil, A. Rare Earth Element Preconcentration from Various Primary and Secondary Sources by Polymeric Ion Exchange Resins. *Separation & Purification Reviews* **2021**, *0* (0), 1–16. https://doi.org/10.1080/15422119.2021.1993255.
(27) Salo, M.; Knauf, O.; Mäkinen, J.; Yang, X.; Koukkari, P. Integrated Acid Leaching and Biological Sulfate Reduction of Phosphogypsum for REE Recovery. *Minerals Engineering* **2020**, *155*, 106408. https://doi.org/10.1016/j.mineng.2020.106408.
(28) Thompson, V. S.; Gupta, M.; Jin, H.; Vahidi, E.; Yim, M.; Jindra, M. A.; Nguyen, V.; Fujita, Y.; Sutherland, J. W.; Jiao, Y.; Reed, D. W. Techno-Economic and Life Cycle Analysis for Bioleaching Rare-Earth Elements from Waste Materials. *ACS Sustainable Chem. Eng.* **2018**, *6* (2), 1602–1609. https://doi.org/10.1021/acssuschemeng.7b02771.
(29) Habashi, F. The Recovery of the Lanthanides from Phosphate Rock. *Journal of Chemical Technology and Biotechnology. Chemical Technology* **1985**, *35* (1), 5–14. https://doi.org/10.1002/jctb.5040350103.
(30) Jarosiński, A.; Kowalczyk, J.; Mazanek, Cz. Development of the Polish Wasteless Technology of Apatite Phosphogypsum Utilization with Recovery of Rare Earths. *Journal of Alloys and Compounds* **1993**, *200* (1), 147–150. https://doi.org/10.1016/0925-8388(93)90485-6.
(31) Walawalkar, M.; Nichol, C. K.; Azimi, G. Process Investigation of the Acid Leaching of Rare Earth Elements from Phosphogypsum Using HCl, HNO3, and H2SO4. *Hydrometallurgy* **2016**, *166*, 195–204. https://doi.org/10.1016/j.hydromet.2016.06.008.
(32) Cánovas, C. R.; Chapron, S.; Arrachart, G.; Pellet-Rostaing, S. Leaching of Rare Earth Elements (REEs) and Impurities from Phosphogypsum: A Preliminary Insight for Further Recovery of Critical Raw Materials. *Journal of Cleaner Production* **2019**, *219*, 225–235. https://doi.org/10.1016/j.jclepro.2019.02.104.
(33) Moni, S. M.; Mahmud, R.; High, K.; Carbajales-Dale, M. Life Cycle Assessment of Emerging Technologies: A Review. *Journal of Industrial Ecology* **2020**, *24* (1), 52–63. https://doi.org/10.1111/jiec.12965.
(34) Li, Y.; T. Trimmer, J.; Hand, S.; Zhang, X.; G. Chambers, K.; C. Lohman, H. A.; Shi, R.; M. Byrne, D.; M. Cook, S.; S. Guest, J. Quantitative Sustainable Design (QSD) for the Prioritization of Research, Development, and Deployment of Technologies: A Tutorial and Review. *Environmental Science: Water Research & Technology* **2022**. https://doi.org/10.1039/D2EW00431C.
(35) *ISO 14040:2006*; Environmental management: Life cycle assessment; Principles and framework; International Organization for Standardization, 2006. https://www.iso.org/standard/37456.html (accessed 2022-05-27).
(36) *ISO 14044:2006*; Environmental management: Life cycle assessment; Requirements and guidelines; International Organization for Standardization, 2006. https://www.iso.org/standard/38498.html (accessed 2022-05-27).

(37) Wernet, G.; Bauer, C.; Steubing, B.; Reinhard, J.; Moreno-Ruiz, E.; Weidema, B. The Ecoinvent Database Version 3 (Part I): Overview and Methodology. *Int J Life Cycle Assess* **2016**, *21* (9), 1218–1230. https://doi.org/10.1007/s11367-016-1087-8.
(38) Huijbregts, M. A. J.; Steinmann, Z. J. N.; Elshout, P. M. F.; Stam, G.; Verones, F.; Vieira, M.; Zijp, M.; Hollander, A.; van Zelm, R. ReCiPe2016: A Harmonised Life Cycle Impact Assessment Method at Midpoint and Endpoint Level. *Int J Life Cycle Assess* **2017**, *22* (2), 138–147. https://doi.org/10.1007/s11367-016-1246-y.
(39) Seider, W. D.; Seader, J. D.; Lewin, D. R. *Product and Process Design Principles: Synthesis, Analysis and Design*, 3rd ed.; John Wiley & Sons, Ltd, 2009.
(40) Perry, R. H.; Green, D. W. *Perry's Chemical Engineers' Platinum Edition*, 7th ed.; McGraw-Hill, 1999.
(41) Tan, E. C. D.; Talmadge, M.; Dutta, A.; Hensley, J.; Schaidle, J.; Biddy, M.; Humbird, D.; Snowden-Swan, L. J.; Ross, J.; Sexton, D.; Yap, R.; Lukas, J. *Process Design and Economics for the Conversion of Lignocellulosic Biomass to Hydrocarbons via Indirect Liquefaction. Thermochemical Research Pathway to High-Octane Gasoline Blendstock Through Methanol/Dimethyl Ether Intermediates*; NREL/TP-5100-62402; National Renewable Energy Lab. (NREL), Golden, CO (United States), 2015. https://doi.org/10.2172/1215006.
(42) Mutel, C. Brightway: An Open Source Framework for Life Cycle Assessment. *Journal of Open Source Software* **2017**, *2* (12), 236. https://doi.org/doi:10.21105/joss.00236.
(43) Cortes-Peña, Y.; Kumar, D.; Singh, V.; Guest, J. S. BioSTEAM: A Fast and Flexible Platform for the Design, Simulation, and Techno-Economic Analysis of Biorefineries under Uncertainty. *ACS Sustainable Chem. Eng.* **2020**, *8* (8), 3302–3310. https://doi.org/10.1021/acssuschemeng.9b07040.
(44) Shi, R.; Guest, J. S. BioSTEAM-LCA: An Integrated Modeling Framework for Agile Life Cycle Assessment of Biorefineries under Uncertainty. *ACS Sustainable Chem. Eng.* **2020**, *8* (51), 18903–18914. https://doi.org/10.1021/acssuschemeng.0c05998.
(45) Li, Y.; Zhang, X.; Morgan, V. L.; Lohman, H. A. C.; Rowles, L. S.; Mittal, S.; Kogler, A.; Cusick, R. D.; Tarpeh, W. A.; Guest, J. S. QSDsan: An Integrated Platform for Quantitative Sustainable Design of Sanitation and Resource Recovery Systems. *Environ. Sci.: Water Res. Technol.* **2022**. https://doi.org/10.1039/D2EW00455K.
(46) Groen, E. A.; Bokkers, E. A. M.; Heijungs, R.; de Boer, I. J. M. Methods for Global Sensitivity Analysis in Life Cycle Assessment. *Int J Life Cycle Assess* **2017**, *22* (7), 1125–1137. https://doi.org/10.1007/s11367-016-1217-3.
(47) El Batouti, M.; Al-Harby, N. F.; Elewa, M. M. A Review on Promising Membrane Technology Approaches for Heavy Metal Removal from Water and Wastewater to Solve Water Crisis. *Water* **2021**, *13* (22), 3241. https://doi.org/10.3390/w13223241.
(48) Qalyoubi, L.; Al-Othman, A.; Al-Asheh, S. Recent Progress and Challenges of Adsorptive Membranes for the Removal of Pollutants from Wastewater. Part II: Environmental Applications. *Case Studies in Chemical and Environmental Engineering* **2021**, *3*, 100102. https://doi.org/10.1016/j.cscee.2021.100102.
(49) Hammas-Nasri, I.; Elgharbi, S.; Ferhi, M.; Horchani-Naifer, K.; Férid, M. Investigation of Phosphogypsum Valorization by the Integration of the Merseburg Method. *New J. Chem.* **2020**, *44* (19), 8010–8017. https://doi.org/10.1039/D0NJ00387E.
(50) Buchner, G. A.; Zimmermann, A. W.; Hohgräve, A. E.; Schomäcker, R. Techno-Economic Assessment Framework for the Chemical Industry—Based on Technology Readiness Levels. *Ind. Eng. Chem. Res.* **2018**, *57* (25), 8502–8517. https://doi.org/10.1021/acs.iecr.8b01248.
(51) Lee, J. C. K.; Wen, Z. Pathways for Greening the Supply of Rare Earth Elements in China. *Nat Sustain* **2018**, *1* (10), 598–605. https://doi.org/10.1038/s41893-018-0154-5.
(52) Bailey, G.; Joyce, P. J.; Schrijvers, D.; Schulze, R.; Sylvestre, A. M.; Sprecher, B.; Vahidi, E.; Dewulf, W.; Van Acker, K. Review and New Life Cycle Assessment for Rare Earth Production from Bastnäsite, Ion Adsorption Clays and Lateritic Monazite. *Resources, Conservation and Recycling* **2020**, *155*, 104675. https://doi.org/10.1016/j.resconrec.2019.104675.
(53) Bakas, I.; Hauschild, M. Z.; Astrup, T. F.; Rosenbaum, R. K. Preparing the Ground for an Operational Handling of Long-Term Emissions in LCA. *Int J Life Cycle Assess* **2015**, *20* (10), 1444–1455. https://doi.org/10.1007/s11367-015-0941-4.
(54) Frischknecht, R.; Jungbluth, N.; Althaus, H.-J.; Bauer, C.; Doka, G.; Dones, R.; Frischnecht, R.; Hischier, R.; Hellweg, S.; Humbert, S.; Jungbluth, N.; Köllner, T.; Loerincik, Y.; Margni, M.; Nemecek, T. *Implementation of Life Cycle Impact Assessment Methods*; ecoinvent report No. 3, v2.0; Swiss Centre for Life Cycle Inventories: Dübendorf, 2007. https://esu-services.ch/fileadmin/download/publicLCI/03_LCIA-Implementation.pdf (accessed 2025-02-13).

(55) Paulillo, A.; Clift, R.; Dodds, J.; Milliken, A.; Palethorpe, S. J.; Lettieri, P. Radiological Impact Assessment Approaches for Life Cycle Assessment: A Review and Possible Ways Forward. *Environ. Rev.* **2018**, *26* (3), 239–254. https://doi.org/10.1139/er-2018-0004.

(56) Boulay, A.-M.; Bare, J.; Benini, L.; Berger, M.; Lathuillière, M. J.; Manzardo, A.; Margni, M.; Motoshita, M.; Núñez, M.; Pastor, A. V.; Ridoutt, B.; Oki, T.; Worbe, S.; Pfister, S. The WULCA Consensus Characterization Model for Water Scarcity Footprints: Assessing Impacts of Water Consumption Based on Available Water Remaining (AWARE). *Int J Life Cycle Assess* **2018**, *23* (2), 368–378. https://doi.org/10.1007/s11367-017-1333-8.

**Supporting Information**

**Advancing the Economic and Environmental Sustainability of Rare Earth Element Recovery from Phosphogypsum**

Authors: Adam Smerigan, Rui Shi*

[a]Department of Chemical Engineering, The Pennsylvania State University, University Park, PA, USA

[b]Institute of Energy and the Environment, The Pennsylvania State University, University Park, PA, USA

*rms6987@psu.edu

# Table of Contents

# S1. Process Design Details

## S1.1. General Details

*A low technological readiness level (TRL) system analysis*

Most of the technologies used in this process design are TRL 3 and below. Therefore, results are highly uncertain. Further, the models are not always built upon first principles due to limited data availability. Therefore, identifying specific targets for technologies should be avoided. However, this system analysis is very useful for establishing quantitative evidence for the feasibility of REE recovery from PG. In addition, this system-level understanding can be used to identify the major impacts and contributions to the sustainability of the process as well as general areas for technology improvement. As succinctly stated by statistician George Box, "All models are wrong, but some are useful".

*A simplified one-component REE model making multiple REO products*

At the start of the process, the mixture of REEs leached from the PG were modeled as a single component, neodymium (selected due to Nd being the most abundant REE within PG while also being representative of light and heavy REEs). After the selective REE separation step, a variety of individual REEs are obtained (each REE still modeled as neodymium). To account for having to refine individual REEs, the process is split into a number of parallel paths to purify each REE into its final REO product. Considering multiple parallel process streams estimates the additional cost due to the loss of economies of scale. For simplicity, the mass flow rate of each individual REE stream was modeled as the total mass flow rate of REEs/number of REEs recovered.

## S1.2. Feedstock: Transportation and Capacity

No transportation costs or environmental impacts from transportation were considered in this study. PG is assumed to be available free of charge at the volume required for the process operation.

To identify a reasonable production capacity, we considered the market size of the product and feedstock. Approximately 254 million metric tons per year of PG is produced globally and is likely to increase into the future. The United States produces ~100 million metric tons per year of the waste.[1] Within the United States, Florida has the largest volume of PG waste being stored in stacks. We assumed ~50% of the US production of PG is in the Florida region. Therefore, the capacity of the plant was estimated to be ~50 million metric tons of PG per year or 5.7 million kg of PG per hour. We rounded this number down to 1 million kg/hr for ease of calculation and analysis. This corresponds to ~43,800 metric tons per year of REOs. The global production of REEs in 2021 was 280,000 metric tons of REOs. Therefore, the production rate of REEs from 1 million kg/hr of PG that was used here is practical considering current market size.[2]

## S1.3. Equipment design details and assumptions

### S1.3.1. Leaching

A conveyor was used to feed the PG from on-site storage to the leaching unit. The capital and operating costs were estimated assuming a width of 1 meter for the belt.[3] The leaching unit was modeled as a continuous countercurrent system. Each stage in the leaching unit was considered a gravity thickener to estimate the capital cost of the unit. Centrifugal pumps with motors were also modeled for the underflow. We used gravity to move fluid from the overflow to downstream operations. A storage tank with a conical roof was priced for holding a week's worth of lixiviant. The following unit operations are included in this section: leaching tanks, PG conveyor, underflow pumps, and a lixiviant storage tank. Figure S8 shows a flow diagram of the industrial-scale leaching unit with some variable definitions for interpreting key assumptions. Table S1 highlights some of the key assumptions when modeling the industrial-scale leaching operation.

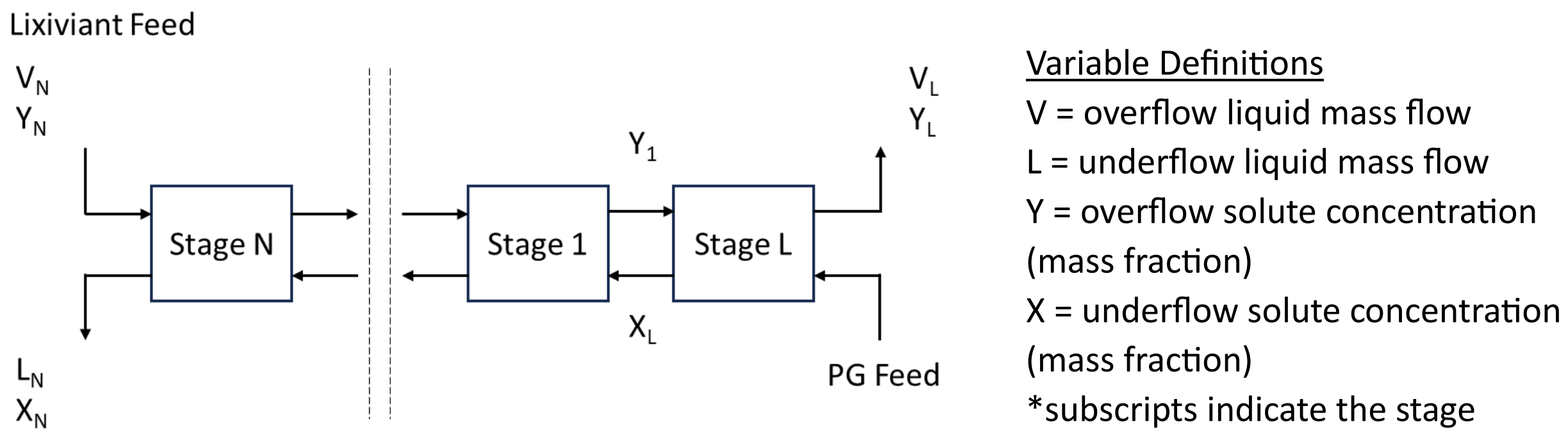

Figure S8: Flow diagram of the continuous countercurrent leaching unit.

Table S1: Assumptions for the leaching unit.

| Assumption | Further Comments |
|---|---|
| Ideal stages (all solute is dissolved in the solvent and the solvent is unchanged) | There are numerous conditions for this assumption to be true: 1) Entering solid solutes can be completely dissolved in the liquid for the stage, 2) Composition of each stage is uniform throughout (well mixed), 3) REE concentration is equal in overflow and underflow (X = Y), 4) Solute is not adsorbed on the surface of the inert solid, 5) The mass ratio of solvent to solids is constant throughout (L constant), 6) Overflows contain no solids. This assumption is used for the mass balance, but the actual proportion of solids in the overflow is modeled as 30% of the total solids, 7) Solvent does not vaporize, get adsorbed, or crystallize |
| Batch, lab-scale data are accurate for a continuous full-scale system | It is likely that the industrial-scale system would have better mass transfer than the experimental results and achieve higher efficiency than is modelled here. |
| Nth-stage underflow REE content is negligible when converging recycle stream | $X_N$ = 0.0001 kg REE/kg solvent is negligible and including this value would only increase the difficult of converging the recycle stream. |
| No interaction between the parameters that were varied individually in the lab scale study[4] when calculating the overall REE leaching efficiency of the system | Realistically, there could be synergistic effects when changing multiple leaching parameters at a time that are not captured here due to lack of data. |
| Actual stage efficiency will be 90% of the theoretical stage efficiency | This inefficiency is used to calculate the real number of stages which affects the capital cost. |
| Aspect ratio of tanks is diameter/height = 6. | Aspect ratio affects the size factor for costing this equipment. |
| PG enters as an anhydrous solid | PG can contain water and acid. However, any liquid content in the PG is compensated by reducing the process water feed to leaching. Therefore, there should be no difference in results based on where this water enters the system, other than for quantifying water reclamation potential. |
| The gypsum will have most of the radionuclides leached during leaching. | After filtration, this gypsum will be saleable to the construction industry.[4] |

## S1.3.2. Filtering

All filters in the process (F1, F2, F3, F4) are modeled as rotary drum vacuum filters. Filtration rates were taken as representative numbers for coarse and fine solids due to the lack of experimental data.[3] These filtration rates were used to calculate the size factor for the filters. The energy required to maintain the vacuum as well as the air flow rate through the solids was taken from a range of numbers given for vacuum filtration of solids.[5] We used the high end of the range of air flows for coarse solids (high energy consumption) and the low end of the range for fine solids (low energy consumption). The air flow through the solids was used to calculate the size factor for the liquid ring vacuum pumps to get the capital cost of the equipment.

Table S2: Assumptions for the rotary filter units.

| Assumption | Further Comments |
|---|---|
| Rotary drum vacuum filters yield a completely dried product | |
| 100% of the solids are filtered | |
| Gypsum is a 'coarse' solid[3] | Influences the amount of energy consumed |
| Heavy metal (REEs, U, and Th) complexes and oxides are 'fine' solids[3] | Influences the amount of energy consumed |
| Cost equations are valid for the full range of plant capacities and process scenarios | The maximum size of these units is used to calculate the number of parallel units required, which should allow the system costs to scale correctly with capacity. |

## S1.3.3. Precipitation

Precipitation steps (P1, P2, P3) were modeled as gravity sedimentation tanks. The chemicals were mixed in the feed and the resulting precipitates were allowed to settle and exit out of the bottom of the tank. P2 was modeled as several parallel tanks for each individual REE. The amount of oxalic acid fed was calculated based on industry best-practices.[6] P1 was modeled as a single tank and the oxalic acid feed rate was calculated similarly to P2. However, additional ions in the leachate will also complex with oxalic acid requiring a higher oxalic acid flow rate. The most abundant contaminant is calcium, which was modeled as the concentration at saturation from calcium sulfate. From this concentration, the amount of additional oxalic acid required was calculated. P3 was modeled as a single unit that neutralizes the wastewater through the addition of sodium hydroxide and precipitates the radionuclides by adding trisodium phosphate. Visual Minteq simulations indicated that both uranium and thorium precipitated at ~pH 9 with the addition of excess phosphates in the solution. Uncertainty in all the precipitant feed rates was considered in the uncertainty analysis.

Table S3: Assumptions for the precipitation units.

| Assumption | Further Comments |
|---|---|
| 100% precipitation of target ions for respective precipitation tanks (REEs, U, and Th) | This assumption was based on literature values.[6,7] We then verified using VisualMinteq simulations. Further, we accounted for this uncertainty by assuming that REEs less than 1 wt% of the total REE content were too dilute to be recovered by this system. |
| P1 and P2: no precipitation of other metal ions with the REEs | Oxalic acid complexes with REEs are more favorable and less soluble compared to other metals in PG. This property allows for REEs to be precipitated and for U/Th to remain in solution for subsequent downstream treatment. This difference in solubility was confirmed using VisualMinteq simulations. |
| All precipitates will settle and leave in the underflow (no solids leaving in the overflow) | We assumed that 2 hrs is enough time to settle all of the solids. |
| Precipitation underflow solids contents were assumed based on total mass flow rates of solids and liquids in the unit operation | This assumption was verified by ensuring the feed solid content, underflow solid content, unit area ($m^2/(mt/day)$), and overflow rate ($m^3/m^2/hr$) are in the ranges of existing gravity sedimentation units.[8] |

## S1.3.4. Selective REE Separation

In contrast to the current paradigm of inefficient organic phase selective separations of REEs (namely solvent extraction), we explore the potential use of aqueous separation methods for REE separations. Though still at the proof-of-concept stage, these aqueous separations have the potential to increase efficiency using novel and non-toxic aqueous phase ligands exhibiting highly stratified binding affinity between adjacent REEs. This stratified affinity (selectivity) can reduce toxic chemical consumption, energy consumption, and cost leading to overall more sustainable operations. However current research lacks system-level understanding on the required performance of such a technology.

Here we modeled the selective REE separation unit as an ion exchange column utilizing a biobased solid adsorbent due to their good lab-scale performance[9,10], high research interest, and adequate data availability. As a specific case study, we modeled the biobased adsorbent using literature data for the protein Lanmodulin. However, we kept the model very generalizable to grant insight into the feasibility of other aqueous ligands (e.g., peptides) that may have different selectivity and cost. Due to the low technological readiness (TRL < 3) of these biobased adsorbents, this model has many limitations. This model does not rigorously model adsorption (due to the lack of kinetic and thermodynamic data) and is limited to estimating costs using only the maximum adsorption capacity and immobilization density. This approach does allow the model to be very simple and flexible, but at the cost of accuracy leading to highly uncertain results. Further, this model does not identify or quantify the regenerant (no cost or environmental impact) and does not consider pumping costs. Therefore, the results from this model should only be used to begin to develop a system-level understanding of the requirements for a theoretical aqueous biobased adsorbent separation for REE recovery. The high uncertainty and lack of a detailed model precludes the ability to identify specific research targets and insights for this technology. Ultimately, we provide quantitative evidence that a biobased adsorbent ***has the potential*** to disrupt the current REE separation paradigm and is worth funding for further investigation. Further conclusions on specific details of the separation are tentative and should be regarded as such. Detailed assumptions and comments on the modeling of this unit is provided below.

Table 4: Base case values of important parameters for the selective separation

| Design Variables - These control the cost of the unit | | | Reference |
|---|---|---|---|
| capacity | 0.00577 | mol/L | [9] |
| immobilization density | 2.47 | mmol/L | [9] |
| cycle time | 4 | hrs | Assumed |
| Biomolecule price | 0.5 | $/g | Estimate[11–13] |
| resin price | 45 | $/L resin | [14] |
| adsorbent lifetime | 10 | years | Assumed |
| plant lifetime | 30 | years | - |
| | | | |
| Simple model based on the assumptions above | | | |
| MW of biomolecule | 11,800 | g/mol | LanM protein |
| Inlet Flow of REE | 12,290 | mol/hr | |
| Resin Needed | 17,040,804 | L | |
| Biomolecule Needed | 496,671,280 | g | |
| Resin Cost | 766,836,191 | $ | |
| Biomolecule Cost | 248,335,640 | $ | |
| Adsorbent Capital Cost | 1,015,171,831 | $ | |
| Replacements | 2.00 | # of | |
| Adsorbent Replacement Cost | 67,678,122 | $/year | |
| | | | |
| purchase cost | 59.57 | $/L adsorbent | |

Equations 1-6 were used to calculate the cost of the unit. Two parallel units were costed to allow for regeneration of the adsorbent material. The adsorbent capital cost is included in the capital expense during the construction period (without including installation cost per the comment above). The adsorbent replacement cost was annualized over the plant lifetime and included as an operating expense.

$$\text{Resin Needed} = \frac{\text{Molar Feed of REE} * \text{Cycle Time} * 2 \text{ Units}}{\text{Capacity}} \quad \text{Eqn 1}$$

$$\text{Resin Cost} = \text{Resin Price} * \text{Resin Needed} \quad \text{Eqn 2}$$

$$\text{Biomolecule Cost} = \text{Biomolecule Price} * \text{Resin Needed} * \text{Immobilization Density} * \text{MW of Biomolecule} \quad \text{Eqn 3}$$

$$\text{Adsorbent Capital Cost} = \text{Resin Cost} + \text{Protein Cost} \quad \text{Eqn 4}$$

$$\text{Replacements} = \frac{\text{Plant Lifetime} - \text{Adsorbent Lifetime}}{\text{Adsorbent Lifetime}} \quad \text{Eqn 5}$$

$$\text{Adsorbent Replacement Cost} = \frac{\text{Adsorbent Capital Cost} * \text{Replacements}}{\text{Plant Lifetime}} \quad \text{Eqn 6}$$

Table 5: Assumptions for the selective separation unit.

| Assumption | Further Comments |
|---|---|
| The solid-state adsorbent is highly selective (very high separation factors between adjacent REEs) for individual REEs and will not require additional stages to create a high purity individual REE product that can replace those currently being sold on the market. | The “REE recovery” technological parameter can be a proxy to account for the uncertainty due to inefficiencies from poor selectivity (poor selectivity would make it difficult to completely separate REEs into individual elements). However, it is not linked explicitly to selectivity at this point due to lack of data availability. Current studies are exploring multiple routes to increase selectivity of the unit.[10] |
| The maximum adsorption capacity reported for LanM on agarose resin[9] is equivalent to the dynamic adsorption capacity. | The dynamic adsorption capacity accounts for when the adsorbent becomes ‘exhausted’ at a specific REE concentration and feed conditions. |
| That the capacity determined experimentally[9] is the same when exposed to the simulated conditions of this study. | The pH for this simulation and the experiment are ~6 and 5, respectively. The pH in this study is below when significant metal hydrolysis would occur (pH ~7) and above when desorption would occur (pH ~3)[9]. The LanM protein binding site in EF-hand 1 is destabilized at pH < 5.[15] Therefore, we might expect capacity to be similar or higher at ~pH 6.<br>The REE concentration for this simulation and the experiment are ~1 M and 0.2 mM, respectively. We would expect higher REE concentrations to yield higher capacity by shifting the equilibrium towards the bound complex. |
| A scaled-up system will perform similarly to the lab-scale setup. | In a full-scale system, changes in unit size will change the inlet flow rate, capacity, and productivity of the unit. Due to the lack of kinetic and thermodynamic data, a detailed model predicting the performance of a large-scale application of this technology is impossible. Further, we are unable to recommend specific configurations to optimize the process design. |
| The cost of the unit is the sum of the cost of a specialized adsorption resin and the cost of the bio-adsorbent (e.g., protein) attached to this resin at the specified immobilization density from experiment. | |
| The installed cost of the separation will be negligible compared to the cost of the resin | To estimate the installed cost, an industrial average factor (~4, based on known purchased and installed costs for chemical plants) can typically be used as an estimate. However, since the cost of the resins can be very large in this model, applying this installation factor greatly overestimates the installed cost. In the absence of actual installation costs for similar systems, we think ignoring the installation cost of this system leads to more accurate and meaningful results. As opposed to multiplying by a factor, we considered adding an ‘average’ installation cost for an ion exchange column. However, adding another uncertain number to an already highly uncertain model seemed to only add more complexity for a marginal improvement to accuracy. |

## S1.3.5. Fired Heaters

The fired heaters in this process are used to burn off the organic compounds from REE-oxalate coordination complexes that precipitated from solution. The temperature of the heater and the time the REEs are heated[6] are used to estimate the heat duty for each heater. A 1:1 volume ratio of air to solids was assumed along with 80% efficiency of heating. The heat duty was modeled as the heat required to change the temperature of the air, solids, and water within the solids from room temperature to the specified temperature (including heat of vaporization for water and the heat from combusting the oxalic acid). Natural gas is used to satisfy the heating requirements.

### S1.3.6. Wastewater Treatment

The wastewater treatment plant (WWT) is modeled as a primary, secondary, and tertiary plant that takes in 3 wastewater streams and has one clean water outlet stream. The capital cost[3] of a generalized WWT plant and its operating cost[16] are calculated based on the water flowrate.

# S2. Technological and System Parameters

In the following three tables, we list the key system and technological parameters used in the system model. We have broken these parameters down into system parameters (Table S6), technological parameters that are not included in the global uncertainty and sensitivity analysis (Table S7), and technological parameters that are included in the global uncertainty and sensitivity analysis (Table S8). Parameters not included in the uncertainty are physically constrained (where other values would exceed the bounds of reasonable operation of that equipment). Parameters included in the uncertainty analysis are bound with an uncertainty range based either on literature evidence or engineering judgement. Uniform distributions are used for parameters that are equally likely to have any value within the lower and upper bounds, represented as (lower, upper) in Table S8. Triangle distributions are used for parameters that have a known value (base) but may vary around this value with decreasing likelihood the further you go from the base value (closer to the lower and upper bounds), represented as (lower, base, upper) in Table S8.

Table S6: System parameters defined prior to simulation that outline system scale and feedstock composition.

| Decision Variable | Units | Values[a] | Reference |
|---|---|---|---|
| Capacity (PG flow rate) | M kg PG/hr | 0.1-2 | - |
| REE Content of PG | wt % | 0.01-1 | [17] |
| Number of recoverable REEs | - | 9 | See assumption 1 in Table S3 |
| Radionuclide content in the PG | wt % | 0.003159 | [4] |

Table S7: Technological parameters not included in the uncertainty and sensitivity analysis.

| Unit Operation | Unit ID | Decision Variable | Units | Values | Reference |
|---|---|---|---|---|---|
| Precipitation | P1 | Underflow Solids Concentration | wt % solids | 5 | [8] |
| | P2 | Underflow Solids Concentration | wt % solids | 50 | [8] |
| | P3 | Underflow Solids Concentration | wt % solids | 5 | [8] |
| Fired Heater | H1 | Temperature | °C | 850 | [6] |
| | H1 | Residence Time | hr | 1.5 | [6] |
| | H2 | Temperature | °C | 850 | [6] |
| | H2 | Residence Time | hr | 1.5 | [6] |
| Vacuum Filter | F3 | Filtration Rate | $kg/hr/m^2$ | 305.15 | [3] |
| | F4 | Filtration Rate | $kg/hr/m^2$ | 305.15 | [3] |
| | F1 | Filtration Rate | $kg/hr/m^2$ | 1220.5 | [3] |
| | F2 | Filtration Rate | $kg/hr/m^2$ | 1220.5 | [3] |

Table S8: Technological parameters used in the global uncertainty and sensitivity analysis.

| Unit Operation | Unit ID | Decision Variable | Units | Distribution | Values[a] | Reference |
|---|---|---|---|---|---|---|
| Leaching | U1 | Time | min | Triangle | 190, 200, 205 | [4] |
| | U1 | Temperature | °C | Triangle | 44.7, 47, 49.4 | [4] |
| | U1 | Acid Concentration | wt % | Triangle | 3.04, 3.2, 3.36 | [4] |
| | U1 | Solvent to Solid Ratio | - | Triangle | 2.61, 2.75, 2.89 | [4] |
| | U1 | Underflow/Overflow Ratio | - | Triangle | 0.16, 0.2, 0.24 | Assumed |
| | U1 | REE content in solvent feed ($Y_N$) | wt % REE | - | 0 | Assumed |
| | U1 | REE content of the outlet underflow | wt % REE | - | 0.001 | Assumed |
| Precipitation | P1 | Oxalic Acid Feed | kg/hr | Triangle | 0.75, 1, 1.25 | [6] |
| | P2 | Oxalic Acid Feed | kg/hr | Triangle | 0.85, 1, 1.15 | [6] |
| | P3 | NaOH Feed | kg/hr | Triangle | 0.75, 1, 1.25 | Calculated |
| | P3 | $Na_3PO_4$ Feed | kg/hr | Triangle | 0.75, 1, 1.25 | Calculated |
| | P1 | Residence Time | min | Uniform | 60, 180 | Assumed |
| | P2 | Residence Time | min | Uniform | 60, 180 | Assumed |
| | P3 | Residence Time | min | Uniform | 60, 180 | Assumed |
| Selective Separation | S1 | REE Recovery | % | Uniform | 90, 100 | Assumed |
| | S1 | Adsorbent Capacity | mol/L adsorbent | Triangle | 0.0005, 0.006, 5 | [9,18] |
| | S1 | Immobilization Density | mmol/L adsorbent bed | Triangle | 0.247, 2.47, 24.7 | [9] |
| | S1 | Cycle Time | hrs | Triangle | 2, 4, 24 | Assumed |
| | S1 | Adsorbent Lifetime | years | Triangle | 1, 10, 20 | Assumed |
| Vacuum Filter | F1 | Vacuum Pump Electricity Consumption | $kW/m^2$ | Triangle | 9.6, 12, 14.4 | [5] |
| | F2 | Vacuum Pump Electricity Consumption | $kW/m^2$ | Triangle | 9.6, 12, 14.4 | [5] |
| | F1 | Air Flow through Filter Cake | $m^3/m^2/min$ | Triangle | 1.2, 1.5, 1.8 | [5] |
| | F2 | Air Flow through Filter Cake | $m^3/m^2/min$ | Triangle | 1.2, 1.5, 1.8 | [5] |

[a]Values can be represented as a single value that is not varied in the uncertainty/sensitivity analysis, a range of values (e.g., 1-4), or a distribution of values (uniform or triangle). Uniform distributions are used for parameters that are equally likely to have any value within the lower and upper bounds, represented as (lower, upper). Triangle distributions are used for parameters that have a known value (base) but may vary around this value with decreasing likelihood the further you go from the base value (closer to the lower and upper bounds), represented as (lower, base, upper).

# S3. Contextual Parameter Details

Contextual parameters are parameters that are controlled by external forces outside the system. Therefore, these parameters are considered separate from the technological values in the sensitivity analysis. To understand total uncertainty, both contextual and technological parameters are used in the Monte Carlo analysis. We have listed the contextual parameters considered in this study in Table S9. Uniform distributions are used for parameters that are equally likely to have any value within the lower and upper bounds, represented as (lower, upper). Triangle distributions are used for parameters that have a known value (base) but may vary around this value with decreasing likelihood the further you go from the base value (closer to the lower and upper bounds), represented as (lower, base, upper).

## S3.1 Contextual Parameter Details

Table S9: All contextual parameters used to model the system.

| | Decision Variable | Units | Distribution | Values | Reference |
|---|---|---|---|---|---|
| TEA | Interest Rate | % | Uniform | 10, 20 | Assumed |
| | Lang Factor | - | Triangle | 3.42, 4.28, 5.14 | [3] |
| | Labor Cost | M $/year | Triangle | 3.33, 4.16, 4.99 | Calculated |
| | Operating Days | days | Triangle | 262, 328, 347 | Assumed |
| | Income Tax Rate | % | Uniform | 0.21, 0.33 | [19] |
| Prices | REO (9 REOs) | $/kg | Triangle | 36.1, 51.5, 67.0 | [4,20–22] |
| | Gypsum | $/g | Triangle | 6.62, 8.28, 9.94 | [23] |
| | Sulfuric Acid | $/kg | Triangle | 0.071, 0.089, 0.107 | [3,24–26] |
| | Oxalic Acid | $/kg | Triangle | 0.684, 0.855, 1.03 | [27,28] |
| | Sodium Hydroxide | $/kg | Triangle | 0.304, 0.380, 0.456 | [3,29–31] |
| | Trisodium Phosphate | $/kg | Triangle | 0.404, 0.505, 0.606 | [32,33] |
| | Nitric Acid | $/kg | Triangle | 0.354, 0.443, 0.532 | [3,34] |
| | Process Water | ₵/kg | Triangle | 0.029, 0.037, 0.044 | [3] |
| | Electricity | ₵/kWh | Uniform | 6.0, 9.8, 16 | [35] |
| | Wastewater Treatment | $/kg | Triangle | 0.25, 0.50, 0.75 | [16] |
| | Biomolecule Price | $/g biomolecule | Triangle | 0.004, 0.5, 10 | [11–13] |
| | Adsorbent Resin | $/L resin | Triangle | 36, 45, 54 | [14] |

## S3.2 REO Basket Price

Even though the process produces pure individual REO products, the system model considers all the REOs as a single stream for simplicity. Therefore, one total REO price is needed ($P_{total}$). This price is calculated by taking the sum of the individual REO prices ($P_{REO}$) multiplied by the fractional abundance of that REO in the PG ($f_{REO}$), as shown in the equation below. REO prices have fluctuated substantially throughout the past 20 years.

$$P_{total} = \sum_{REO} P_{REO} * f_{REO} \quad \text{Eqn 7}$$

Table S10 shows the baseline prices for each REO. In the literature, numerous methods have been employed to estimate uncertainty in the REO market prices. Some studies consider peak values around 2011 after export restrictions from China. Others consider lower values from more recent years. Some studies consider the market price to be an average of these numbers. Others use the minimum and maximum values in a scenario analysis. Since other countries have begun investing in developing domestic sources of REOs, the REO supply will likely be less geopolitically restricted in the future. Therefore, it is unlikely the market will be disrupted as greatly as it was when one country controlled almost the entire supply of REOs. Rather than purely using historical prices as predictions of the future market value, we considered uncertainty in prices based on the REE balance problem[36] (REEs are coproduced but not equally in demand leading abundant REEs to decrease in value). Therefore, we thought it more reasonable to give a wide uncertainty range (±30%) to the REO basket price around current REO prices. Ultimately, REE production processes that have low operational expenditure (OPEX) have been most successful, despite lower basket prices. Systems that prioritize lower OPEX are more likely to survive price fluctuations leading to profitable operations.[37,38]

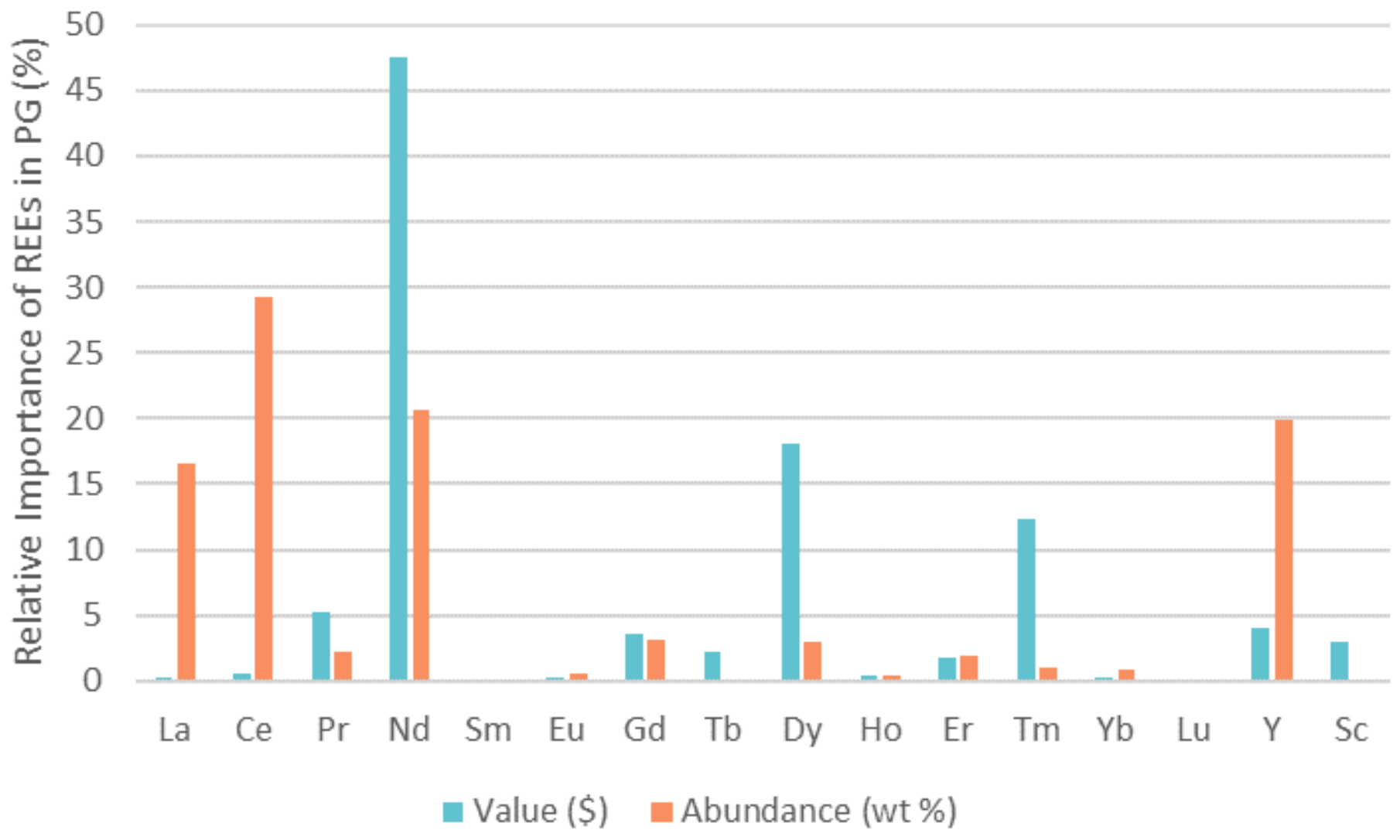

Figure S9: The relative abundance and value of each REE within Florida PG. Neodymium comprises ~50% of the overall value of the REEs within PG. Though La and Ce are relatively abundant, they have relatively low value and will contribute minimally to the bottom line. However, low value REEs will still provide domestic sources of critical materials and increase supply chain security.

Table S10: Individual REO prices used in this study for estimating IRR.

| Rare Earth Oxide | Purity (%) | Price ($/g) | Reference | Date |
|---|---|---|---|---|
| La | 99.5 | 0.001124 | [21] | 07/01/2022 |
| Ce | 99.5 | 0.00119 | [21] | 07/01/2022 |
| Pr | 99.5 | 0.126589 | [21] | 07/01/2022 |
| Nd | 99.5 | 0.126589 | [21] | 07/01/2022 |
| Sm | 99.9 | 0.003109 | [21] | 07/01/2022 |
| Eu | 99.9 | 0.028219 | [21] | 07/01/2022 |
| Gd | 99.5 | 0.062567 | [21] | 07/01/2022 |
| Tb | 99.9 | 1.960128 | [21] | 07/01/2022 |
| Dy | 99.5 | 0.339291 | [21] | 07/01/2022 |
| Ho | 99.5 | 0.05915 | [20] | 08/01/2019 |
| Er | 99.5 | 0.049538 | [21] | 07/01/2022 |
| Tm | 99.99 | 0.659966512 | [22*] | 2022 |
| Yb | 99.99 | 0.01605 | [20] | 08/01/2019 |
| Lu | 99.99 | 0.617 | [20] | 08/01/2019 |
| Y | 99.99 | 0.011 | [21] | 07/01/2022 |
| Sc | 99.99 | 1.05 | [20] | 08/01/2019 |

*Price was extrapolated to estimate the industrially relevant price for 10,000 grams using a power law model.

## S3.3 Biomolecule Price

Biomolecule price is an estimate based on protein and peptide sources at large-scale. Protein literature was more available[11,13] and we confirmed peptide prices predicted in literature[12] with extrapolations from quotes we received. Peptide prices were estimated using a 12 amino acid long sequence. Quotes for producing this peptide were received for 9 different masses varying from 0.004-1000 grams. These quoted prices were further extrapolated to estimate the industrially relevant price for 1 metric tonne of peptide using a power law model. The price estimated from this model ($8.03/g peptide) was closely comparable to other literature estimates (< $12/g peptide[12]). This type of price extrapolation was also completed for $Tm_2O_3$. Both of the regression models for these prices are shown in Figure S10 below.

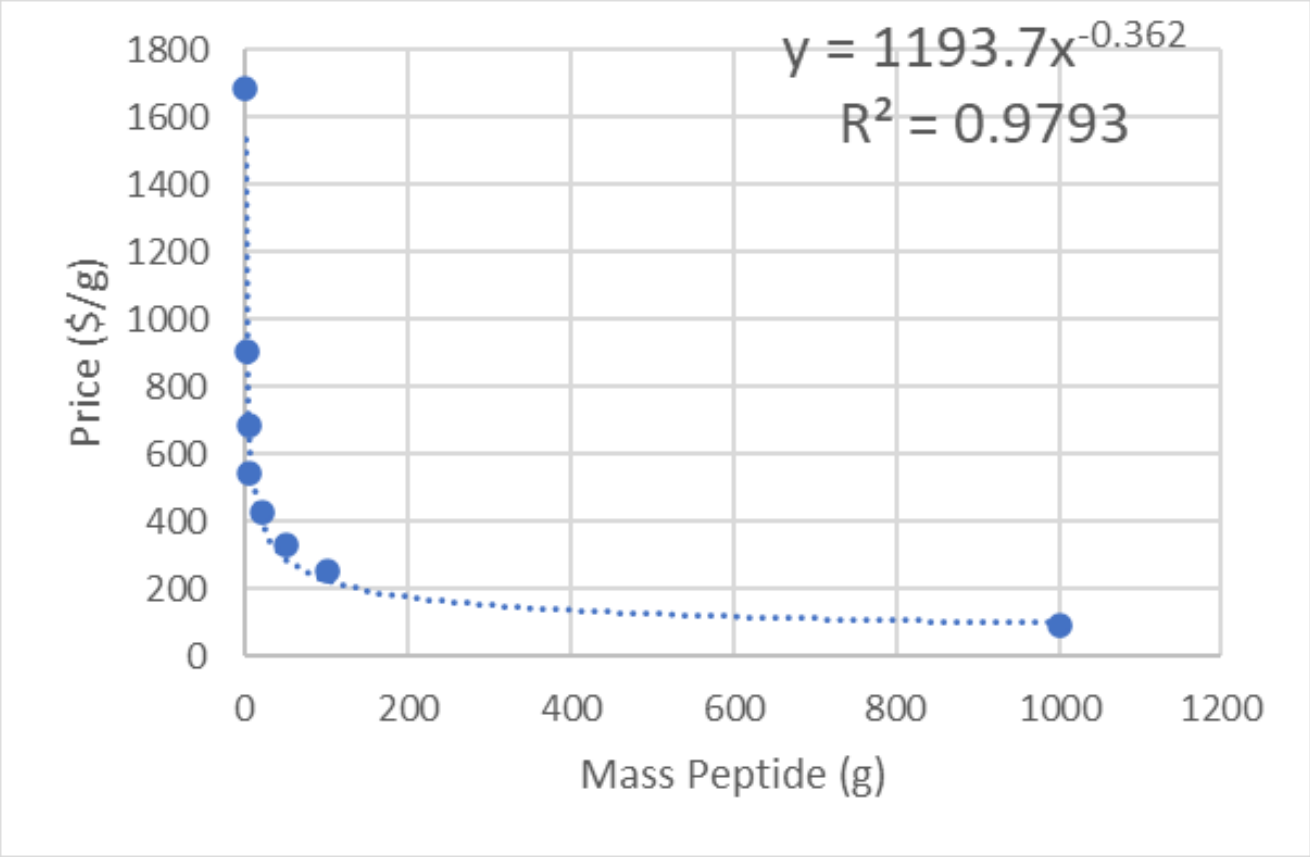

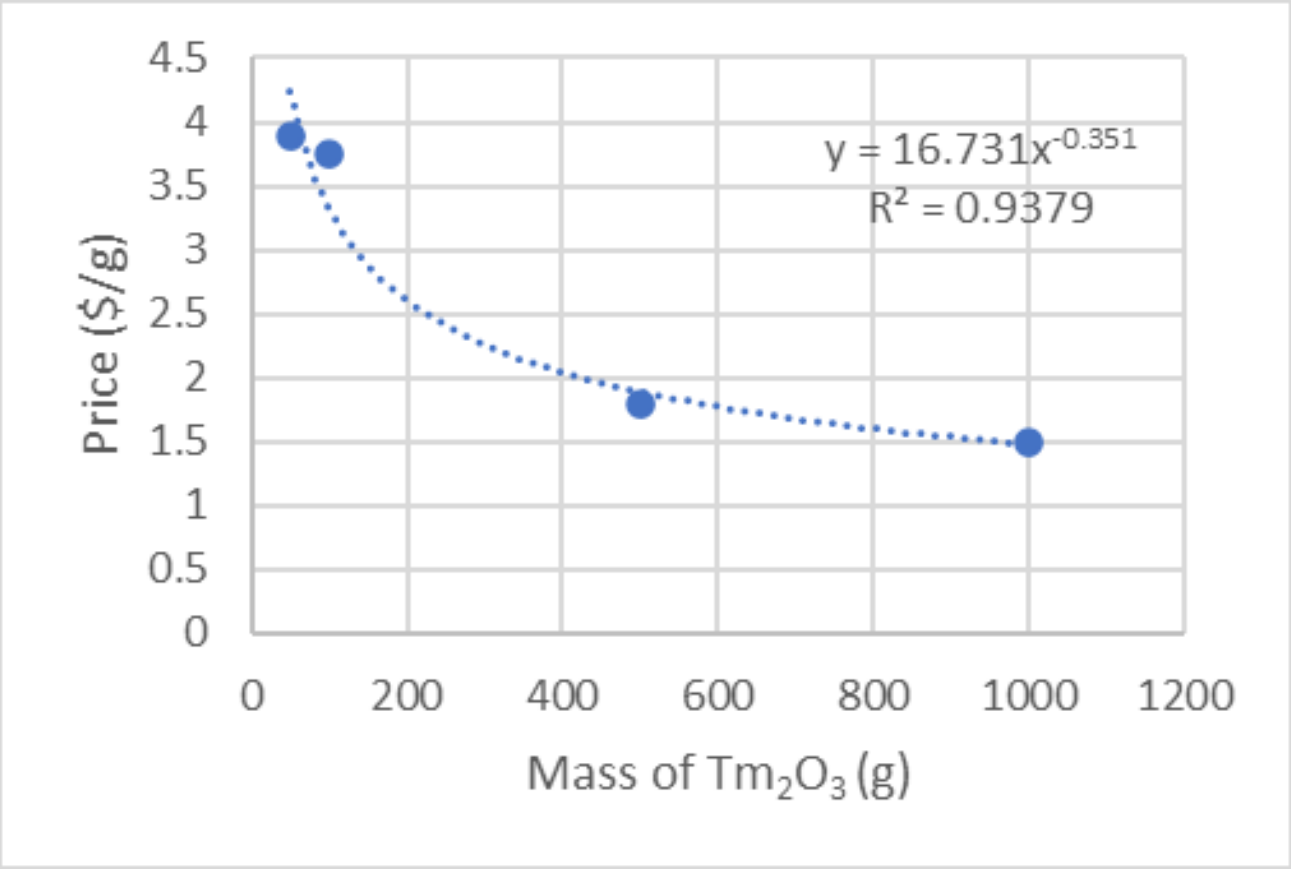

Figure S10: The power law models used to estimate the bulk prices of peptide (left) and thulium oxide (right).

# S4. Optimization of the leaching unit

To get a more realistic perspective on the feasibility of the system, the leaching unit was optimized for highest NPV (the most influential indicator of viability). The optimization was completed using the leaching efficiency data[4] under different leaching temperatures, acid concentrations, residence times, and liquid-to-solid ratios. The full system was simulated for the full range of parameter values in the experimental study to find the global maximum conditions for each parameter. Figure S11 shows the results of the optimization. The values of the parameters at the center of the bright yellow areas (highest profitability) were taken as the optimized values for the baseline simulations in the main text. The black dots represent the experimental local optimal conditions, which are different than the global, system optimal conditions determined here. The original experimental data are adapted here for reference (Figure S12). The maximum reported experimental leaching efficiency was 43% at a sulfuric acid concentration of 5 wt% and 50°C. Increasing leaching time and liquid/solid ratio yielded higher leaching efficiencies. We term the "optimal" conditions as those used to achieve the reported optimal of 43%, which was a leaching time of 120 min and a liquid/solid ratio of 4.0.

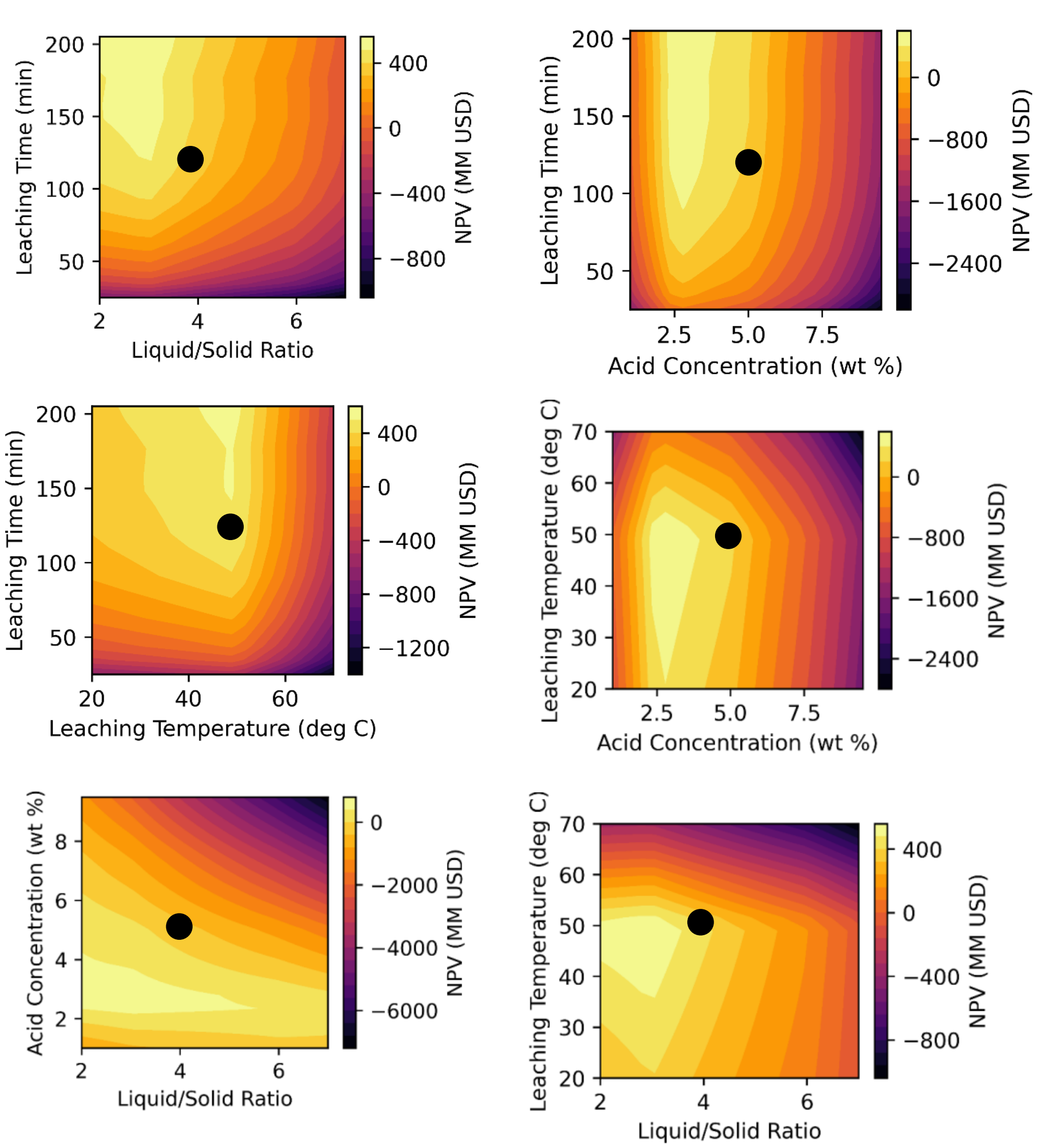

Figure S11: Contour plots showing how the values of key technological parameters were chosen to optimize net present value for the leaching unit. The black dot represents the experimentally determined local optimal conditions, which are different from the system's global optimal conditions (the bright yellow regions).

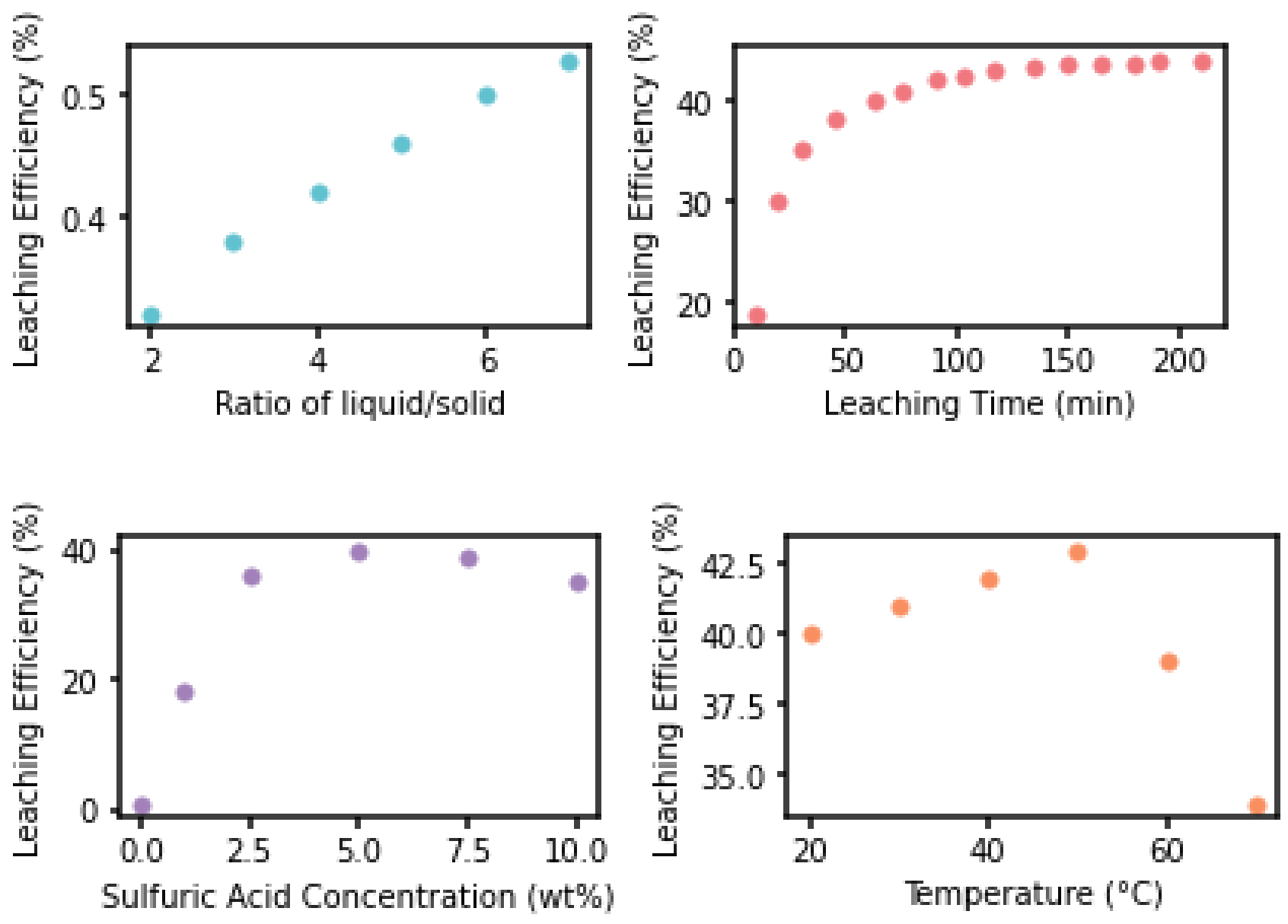

Figure S12: Sulfuric acid leaching data used in the modelling of the leaching unit (adapted from Liang et al.[4]).

# S5. Simulation Results

## S5.1. The effect of the number of samples used in uncertainty and sensitivity analysis

We ran uncertainty analysis using 500, 1000, and 3000 samples to examine how the number of samples influenced the results of the analysis. Further, we replicated the analysis for each number of samples three times to understand how random seeding of the sampling method affected the uncertainty analysis results for the economic indicator NPV15 (Figure S13). Figure S13a shows that the interquartile range is relatively consistent between trials with different numbers of samples. However, the quartile range shows more variation. Additionally, the peak of the distribution of results shifts with the median and the shape of the distribution is slightly dependent on the sampling procedure with the most obvious deformations occurring in the 500 sample results. In Figure S13b, c, and d, the average, median, and standard deviation of NPV15 results are shown. The average varies less than 1% on average. The median shows slightly more variation between 500, 1000, and 3000 samples (3.9%, 2.5%, and 1.4%, respectively). In Figure S13d, the standard increases with number samples as more extreme values are simulated from fringe cases, as shown in Figure S13e. Overall, the effect of increasing the number of samples above 500 is minimal for the number of parameters investigated in this study, especially if you are only concerned with the average result. However, if you are interested in the spread and stability of the results between trials, then additionally sampling can improve consistency in your results, especially for the median value. The improved accuracy must also be weighed against the increased computational expense. Since 3000 simulations only takes about 5 minutes, we chose this number of samples for uncertainty and sensitivity analysis. Studies using a different number of parameters may need fewer or more samples to achieve consistency sampling the design space.

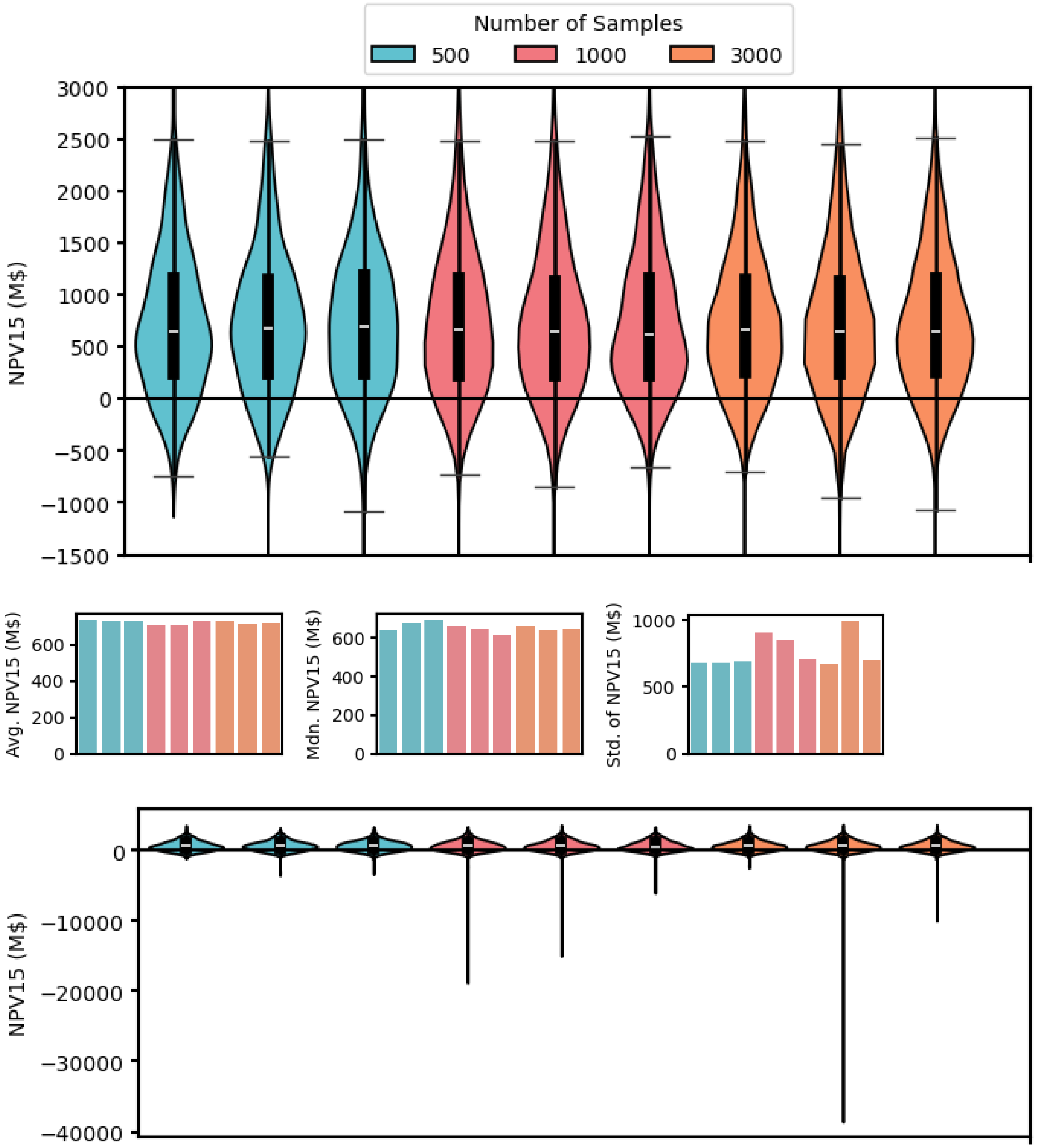

Figure S13: The effect on the number of samples used for the uncertainty and sensitivity analysis shown as (a) violin plots of three trials each for 500, 1000, and 3000 samples. (b)(c)(d) show the average, median, and standard deviation of each of the trials. (e) shows the entire violin plot including outliers.

## S5.2. Uncertainty Analysis Results

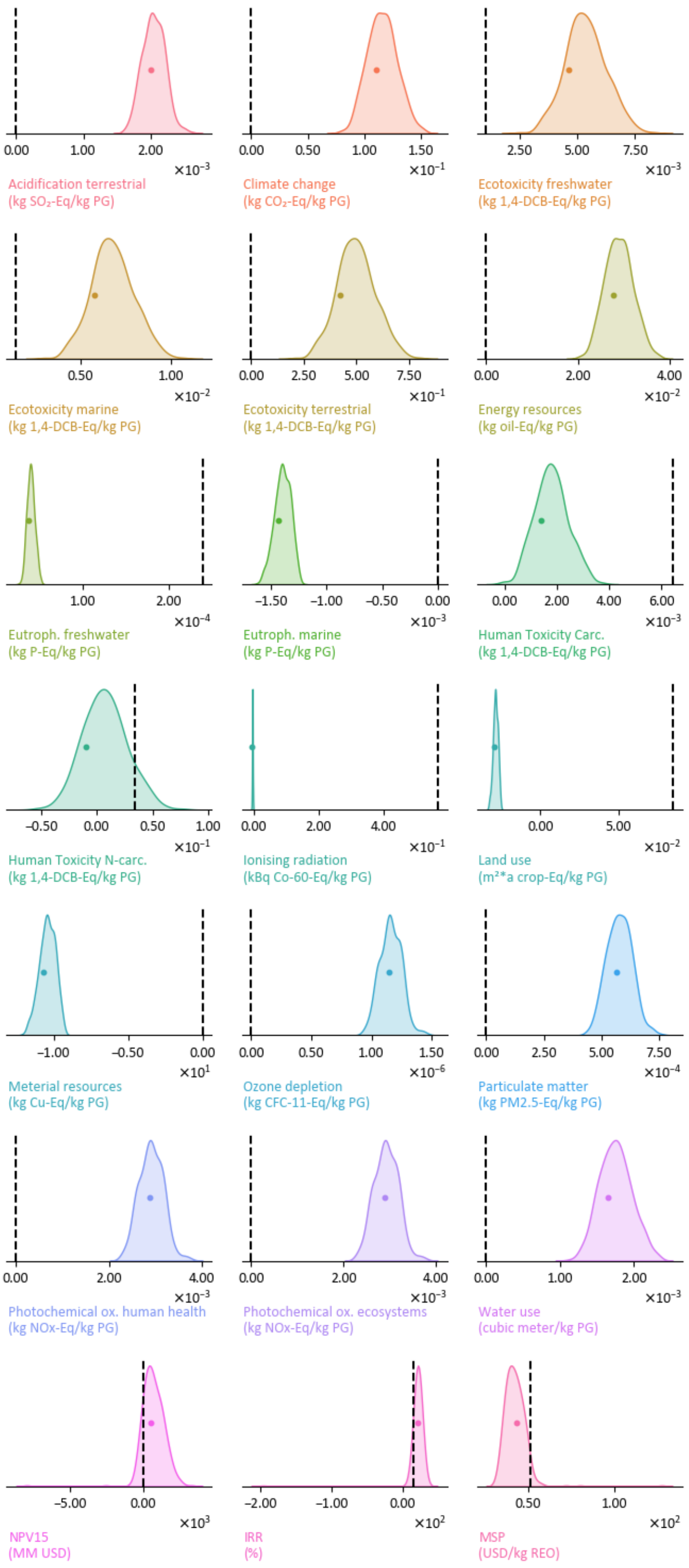

Figure S14: The baseline value (dots) and the probability density (shaded regions) of indicator values for the functional unit of 1 kg of PG produced. Peaks represent a higher probability of an indicator value, while broad distributions represent greater uncertainty. The indicator underlined with a black bar is an economic indicator (net present value at an interest rate of 15%) evaluated by TEA. All other indicators were evaluated by LCA using the ReCiPe 2016 LCIA method. The probability densities were calculated by collecting 3000 samples of the system using Latin Hypercube sampling from defined parameter distributions (Table S8 and Table S9).

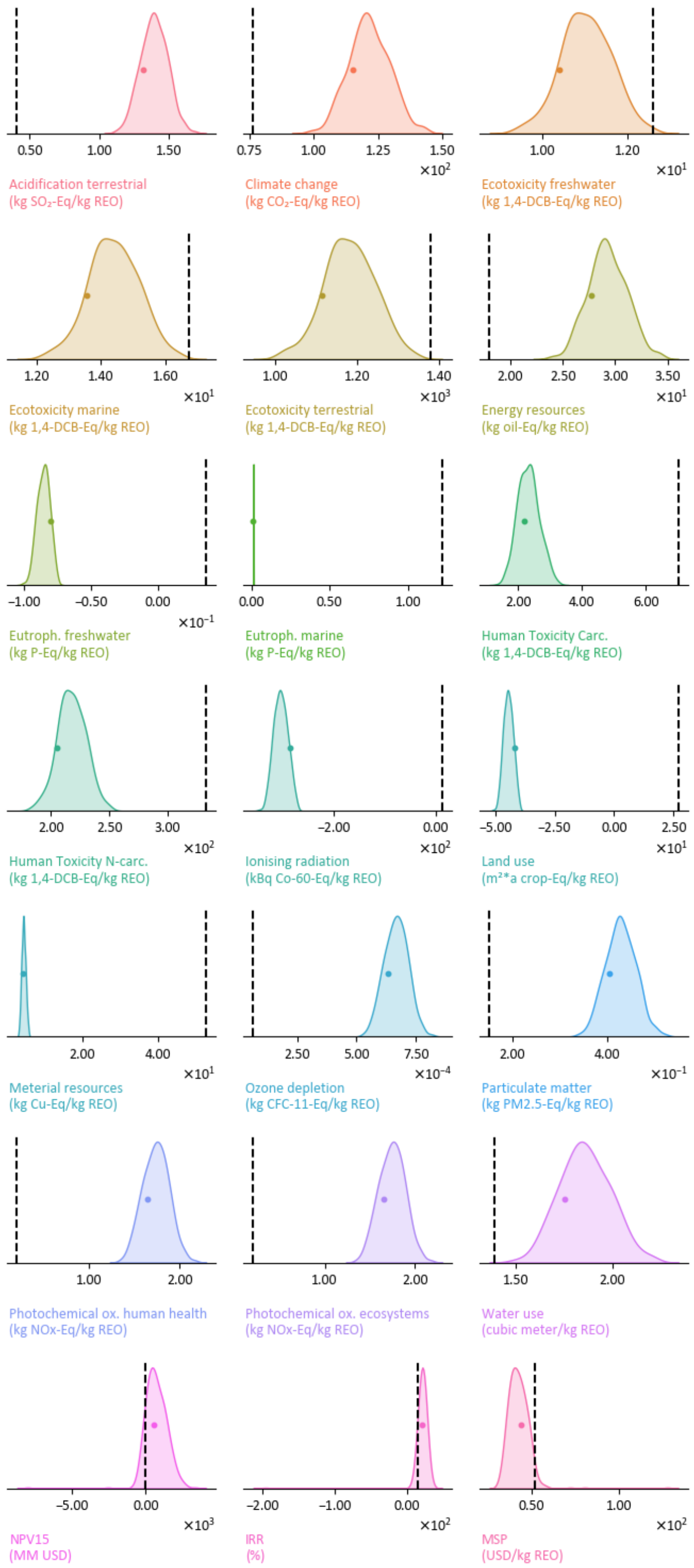

Figure S15: The baseline value (dots) and the probability density (shaded regions) of indicator values for the functional unit of 1 kg of REO produced. Peaks represent a higher probability of an indicator value, while broad distributions represent greater uncertainty. The indicator underlined with a black bar is an economic indicator (net present value at an interest rate of 15%) evaluated by TEA. All other indicators were evaluated by LCA using the ReCiPe 2016 LCIA method. The probability densities were calculated by collecting 3000 samples of the system using Latin Hypercube sampling from defined parameter distributions (Table S8 and Table S9). As more REOs are produced due to higher REE recovery in S1, less PG is being remediated. The credit for PG remediation is larger than the benefit for producing more REO (no LCA credit). Therefore, the overall impact will increase for those categories where there is a large PG credit.

## S5.3. Sensitivity Analysis Results

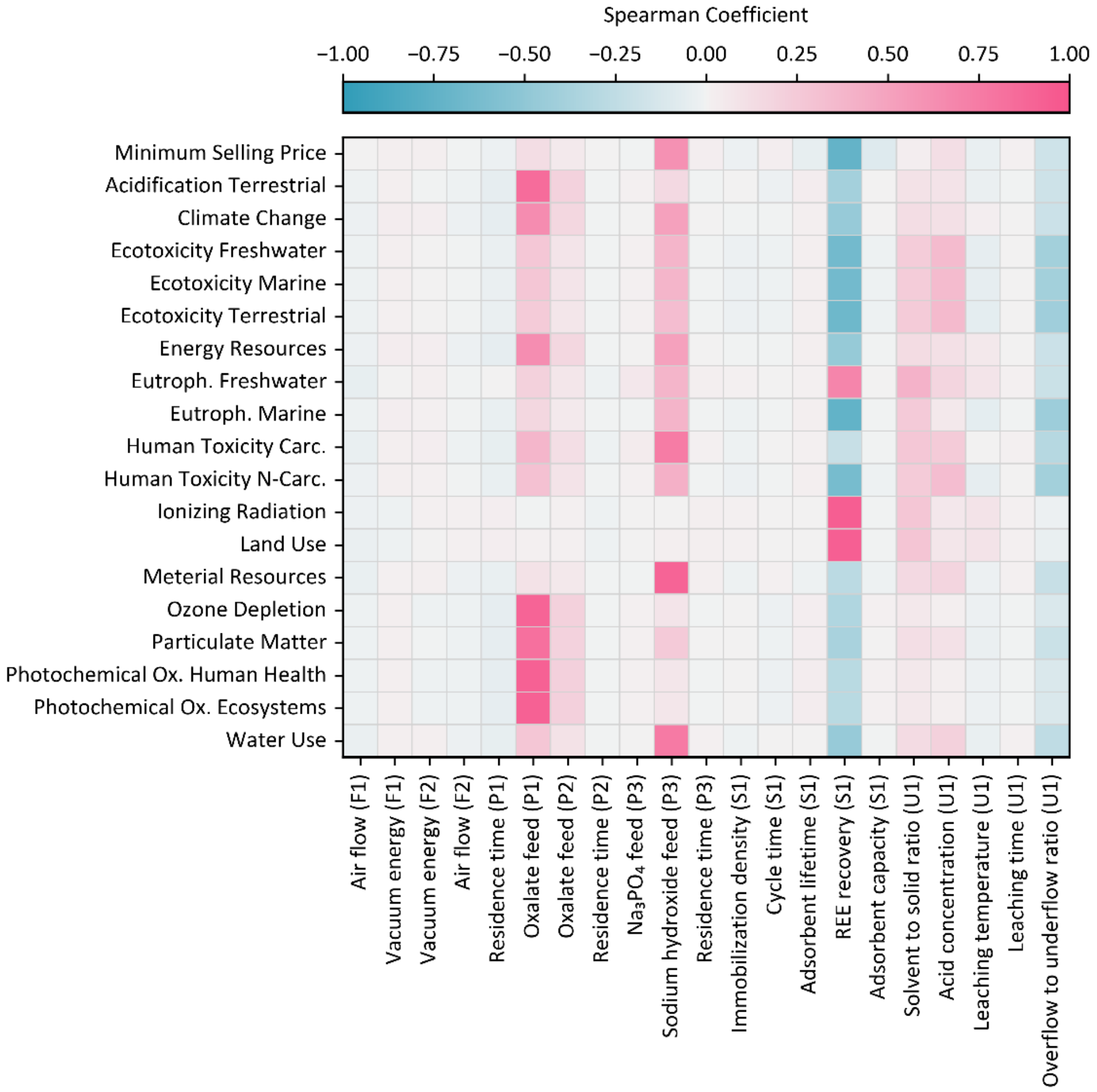

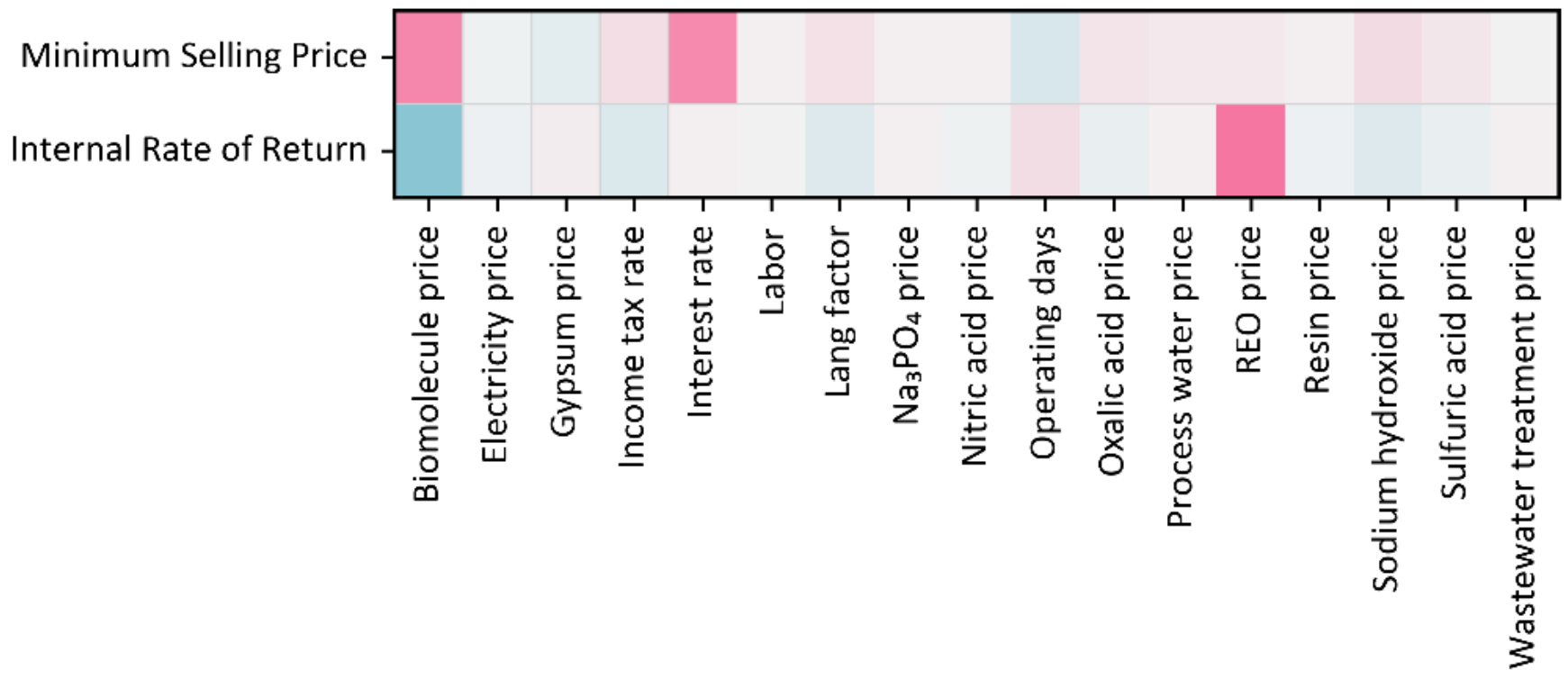

Figure S16: the sensitivity of environmental and economic indicators to **(a)** technological parameters and **(b)** contextual parameters. Blue indicates better performance (reduced environmental impact and higher profitability) with an increase in parameter value, while red indicates inferior performance as the parameter increases. The more vibrant the color, the greater the sensitivity of an indicator (y-axis) to a change in a parameter (x-axis). The indicator above the dotted black line is an economic indicator (evaluated by TEA) and indicators below are environmental indicators (evaluated by LCA for the functional unit of 1 kg of REO produced). Sensitivity was calculated using Spearman rank correlations with Latin Hypercube sampling (3000 samples) of parameter distributions (given in Table S8 and Table S9).

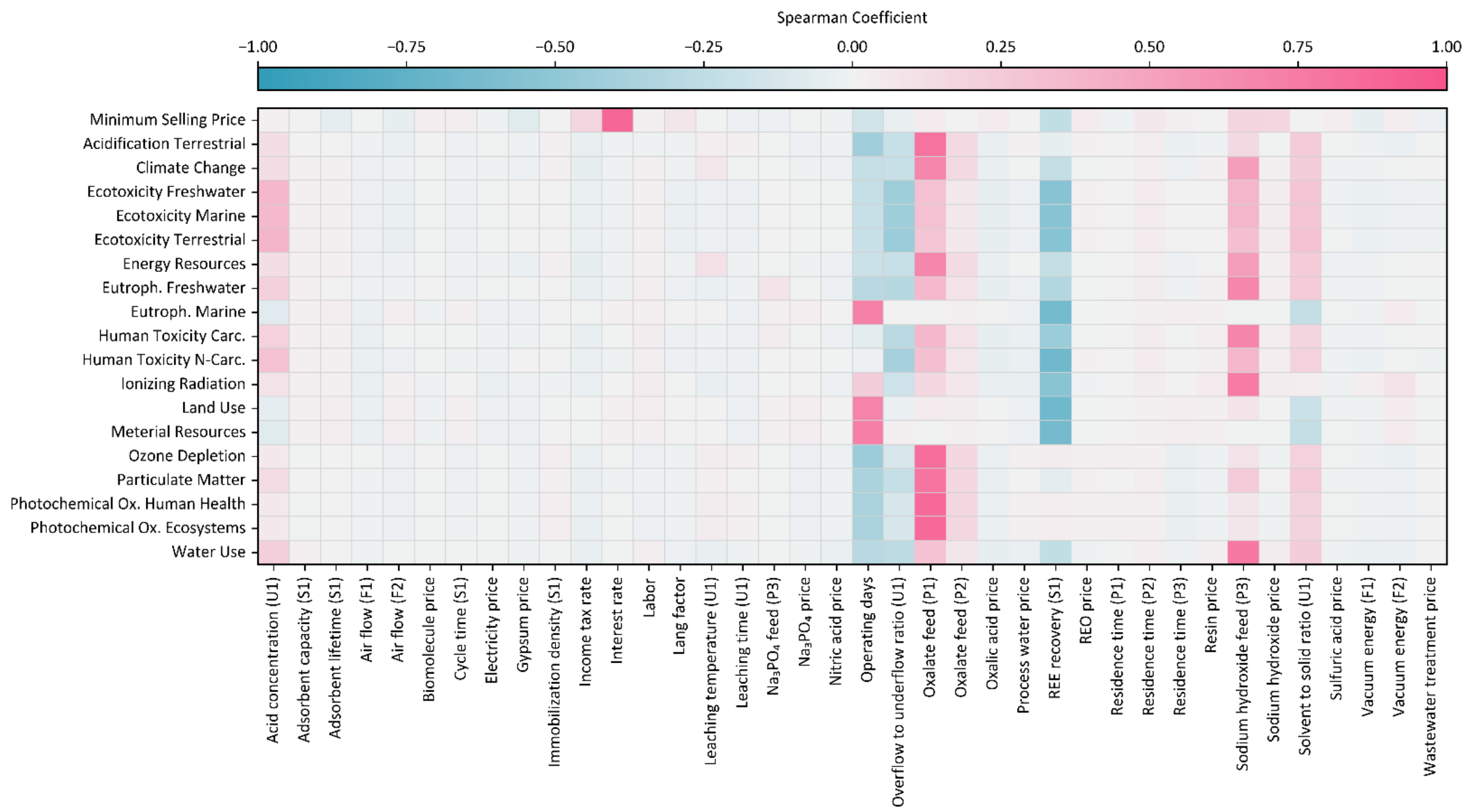

Figure S17: the sensitivity of environmental and economic indicators to **(a)** technological parameters and **(b)** contextual parameters. Blue indicates better performance (reduced environmental impact and higher profitability) with an increase in parameter value, while red indicates inferior performance as the parameter increases. The more vibrant the color, the greater the sensitivity of an indicator (y-axis) to a change in a parameter (x-axis). The indicator above the dotted black line is an economic indicator (evaluated by TEA) and indicators below are environmental indicators (evaluated by LCA for the functional unit of 1 kg of PG produced). Sensitivity was calculated using Spearman rank correlations with Latin Hypercube sampling (3000 samples) of parameter distributions (given in Table S8 and Table S9). Here you can see that combining technological and contextual uncertainty does not lead to changes in conclusions when looking at each individually.

# S6. Life Cycle Inventory

To calculate the life cycle impacts of the system, we developed our background life cycle inventory using ecoinvent v3.9.1. The full list of ecoinvent activities used in the LCA are included in Table S11. Sulfuric acid, process water, heat from natural gas, and electricity was used in conveying the PG feedstock, leaching the PG, and filtering the leachate. Oxalic acid, sodium hydroxide, trisodium phosphate, and electricity were used during precipitation of heavy metals. Nitric acid, process water, and electricity was used to dissolve REEs prior to the selective separation. Heat from natural gas was used during calcination of REE-oxalates. Direct emissions of carbon dioxide and carbon monoxide from calcination are included considering the indirect global warming effect of carbon monoxide. The LCI of the wastewater treatment plant was assumed to be that of an average wastewater treatment system in ecoinvent. Using literature data[39], we created the activity of PG stack treatment (Table S12).

Table S11: List of ecoinvent v3.9.1 activities and flows used in the LCA.

| Inventory Item | Ecoinvent name for activity (geography code) |
|---|---|
| Conventional REO production | market for neodymium oxide (GLO) |
| PG stack treatment | Custom made using flows in Table S12 |
| Gypsum | market for gypsum, mineral (RoW) |
| Sulfuric acid | market for sulfuric acid (RoW) |
| Oxalic acid | market for oxalic acid (Global) |
| Process water | market for water, decarbonized (US) |
| Electricity | market for electricity, medium voltage (US-SERC) |
| Heat from natural gas | heat production, natural gas, at industrial furnace >100kW (CA-QC) |
| Wastewater treatment | market for wastewater, average (RoW) |
| Nitric acid | market for nitric acid, without water, in 50% solution state (RoW)* |
| Sodium hydroxide | market for sodium hydroxide, without water, in 50% solution state (RoW)* |
| Trisodium phosphate | market for trisodium phosphate (GLO) |
| Carbon dioxide | Elementary flow |
| Carbon monoxide | Elementary flow |

*In ecoinvent chemicals products are always expressed in 100% active substance. The inclusion in the activity name of wordings such as "without water, in 50% solution state" simply indicates the most frequent solution state the chemical is found on the market.

Table S12: Ecoinvent flows used to create the custom PG stack activity. The flows used to create the activity were taken from Tsioka and Voudrias[39] and were updated for ecoinvent database v3.9.1 cutoff.[40]

| Flow | Category | Amount | Unit |
|---|---|---|---|
| **Inputs** | | | |
| Occupation, pasture, man made | Resource/land | 153.8 | m2*a |
| Phosphogypsum | Reference flow for waste treatment process | 1 | Mg |
| **Outputs** | | | |
| Cadmium II | Emission to water/ground water | 0.00189 | kg |
| Calcium II | Emission to water/ground water | 12.08 | kg |
| Chromium VI | Emission to water/ground water | 0.00086 | kg |
| Copper ion | Emission to water/ground water | 0.00079 | kg |
| Fluoride | Emission to water/ground water | 118.12 | g |
| Hydrogen Fluoride | Emission to air/low population density | 38.4 | g |
| Particulate Matter, >10 um | Emission to air/low population density | 0.696 | g |
| Phosphate | Emission to water/ground water | 0.726 | kg |
| Radium-226 | Emission to air/low population density | 0.358 | Bq |
| Radium-226 | Emission to water/ground water | 9.27 | kBq |
| Radon-222 | Emission to air/low population density | 3.9206E5 | kBq |
| Silicon Tetrafluoride | Emission to air/low population density | 49.9 | g |
| Sulfate | Emission to water/ground water | 64.59 | kg |
| Uranium-238 | Emission to air/low population density | 0.024 | Bq |
| Zinc II | Emission to water/ground water | 0.00384 | kg |

# S7. References

(1) US EPA, O. *TENORM: Fertilizer and Fertilizer Production Wastes*. https://www.epa.gov/radiation/tenorm-fertilizer-and-fertilizer-production-wastes (accessed 2022-10-26).

(2) *Rare earth mine production worldwide 2021*. Statista. https://www.statista.com/statistics/1187186/global-rare-earths-mine-production/ (accessed 2022-10-26).

(3) Seider, W. D.; Seader, J. D.; Lewin, D. R. *Product and Process Design Principles: Synthesis, Analysis and Design*, 3rd ed.; John Wiley & Sons, Ltd, 2009.

(4) Liang, H.; Zhang, P.; Jin, Z.; DePaoli, D. Rare Earths Recovery and Gypsum Upgrade from Florida Phosphogypsum. *Mining, Metallurgy & Exploration* **2017**, *34* (4), 201–206. https://doi.org/10.19150/mmp.7860.

(5) Henriksson, B. Focus on Separation in the Mining Industry. *Filtration & Separation* **2000**, *37* (7), 26–29. https://doi.org/10.1016/S0015-1882(00)80139-1.

(6) Qi, D. *Hydrometallurgy of Rare Earths: Extraction and Separation*, first.; Elsevier, 2018.

(7) Sadri, F.; Nazari, A. M.; Ghahreman, A. A Review on the Cracking, Baking and Leaching Processes of Rare Earth Element Concentrates. *Journal of Rare Earths* **2017**, *35* (8), 739–752. https://doi.org/10.1016/S1002-0721(17)60971-2.

(8) Perry, R. H.; Green, D. W. *Perry's Chemical Engineers' Handbook*, seventh.; McGraw-Hill, 1999.

(9) Dong, Z.; Mattocks, J. A.; Deblonde, G. J.-P.; Hu, D.; Jiao, Y.; Cotruvo, J. A. Jr.; Park, D. M. Bridging Hydrometallurgy and Biochemistry: A Protein-Based Process for Recovery and Separation of Rare Earth Elements. *ACS Cent. Sci.* **2021**, *7* (11), 1798–1808. https://doi.org/10.1021/acscentsci.1c00724.

(10) Dong, Z.; Mattocks, J. A.; Seidel, J. A.; Cotruvo, J. A.; Park, D. M. Protein-Based Approach for High-Purity Sc, Y, and Grouped Lanthanide Separation. *Separation and Purification Technology* **2024**, *333*, 125919. https://doi.org/10.1016/j.seppur.2023.125919.

(11) *High-Purity Cellulase Enzyme for Sale | Affordable & Global Delivery*. Procurenet Limited. https://procure-net.com/product/cellulase-us35-kg/ (accessed 2024-03-11).

(12) Bray, B. L. Large-Scale Manufacture of Peptide Therapeutics by Chemical Synthesis. *Nat Rev Drug Discov* **2003**, *2* (7), 587–593. https://doi.org/10.1038/nrd1133.

(13) Humbird, D.; Davis, R.; Tao, L.; Kinchin, C.; Hsu, D.; Aden, A.; Schoen, P.; Lukas, J.; Olthof, B.; Worley, M.; Sexton, D.; Dudgeon, D. *Process Design and Economics for Biochemical Conversion of Lignocellulosic Biomass to Ethanol: Dilute-Acid Pretreatment and Enzymatic Hydrolysis of Corn Stover*; NREL/TP-5100-47764; National Renewable Energy Lab. (NREL), Golden, CO (United States), 2011. https://doi.org/10.2172/1013269.

(14) *How Much Does It Cost to Buy, Maintain, and Dispose of Ion Exchange Resins?*. SAMCO Technologies. https://samcotech.com/how-much-does-it-cost-to-buy-maintain-and-dispose-of-ion-exchange-resins/ (accessed 2023-11-08).

(15) Featherston, E. R.; Issertell, E. J.; Cotruvo, J. A. Jr. Probing Lanmodulin's Lanthanide Recognition via Sensitized Luminescence Yields a Platform for Quantification of Terbium in Acid Mine Drainage. *J. Am. Chem. Soc.* **2021**, *143* (35), 14287–14299. https://doi.org/10.1021/jacs.1c06360.

(16) Turton, R.; Shaeiwitz, J.; Bhattacharyya, D.; Whiting, W. *Analysis, Synthesis, and Design of Chemical Processes*, fifth.; Pearson, 2018.

(17) Mukaba, J.-L.; Eze, C. P.; Pereao, O.; Petrik, L. F. Rare Earths' Recovery from Phosphogypsum: An Overview on Direct and Indirect Leaching Techniques. *Minerals* **2021**, *11* (10), 1051. https://doi.org/10.3390/min11101051.

(18) Hérès, X.; Blet, V.; Di Natale, P.; Ouaattou, A.; Mazouz, H.; Dhiba, D.; Cuer, F. Selective Extraction of Rare Earth Elements from Phosphoric Acid by Ion Exchange Resins. *Metals* **2018**, *8* (9), 682. https://doi.org/10.3390/met8090682.

(19) Fritts, J. *State Corporate Income Tax Rates and Brackets for 2023*. Tax Foundation. https://taxfoundation.org/data/all/state/state-corporate-income-tax-rates-brackets-2023/ (accessed 2023-11-08).
(20) *Current prices for rare earths | Institute for Rare Earths and Metals*. Institut für Seltene Erden und strategische Metalle e.V. https://en.institut-seltene-erden.de/aktuelle-preise-von-seltenen-erden/ (accessed 2023-11-08).
(21) *Rare Earth Metals – MineralPrices.com*. https://mineralprices.com/rare-earth-metals/ (accessed 2023-11-08).
(22) *MSE PRO Thulium Oxide (Tm2O3) 99.99% 4N Powder– MSE Supplies LLC*. https://www.msesupplies.com/products/mse-pro-thulium-oxide-tm-sub-2-sub-o-sub-3-sub-99-99-4n-powder?variant=31799741612090%20Accessed%2010%2F11%2F22 (accessed 2023-11-08).
(23) *Mineral Commodity Summaries 2022*; 2022; U.S. Geological Survey, 2022. https://doi.org/10.3133/mcs2022.
(24) *Sulfuric acid Price and Market Analysis - ECHEMI - ECHEMI.com*. ECHEMI. https://www.echemi.com/productsInformation/pid_Rock19440-sulfuric-acid.html (accessed 2023-11-08).
(25) *Sulphuric Acid Prices, News, Monitor | ChemAnalyst*. https://www.chemanalyst.com/Pricing-data/sulphuric-acid-70 (accessed 2023-11-08).
(26) Kulczycka, J.; Kowalski, Z.; Smol, M.; Wirth, H. Evaluation of the Recovery of Rare Earth Elements (REE) from Phosphogypsum Waste – Case Study of the WIZÓW Chemical Plant (Poland). *Journal of Cleaner Production* **2016**, *113*, 345–354. https://doi.org/10.1016/j.jclepro.2015.11.039.
(27) Boor, V.; Frijns, J. E. B. M.; Perez-Gallent, E.; Giling, E.; Laitinen, A. T.; Goetheer, E. L. V.; van den Broeke, L. J. P.; Kortlever, R.; de Jong, W.; Moultos, O. A.; Vlugt, T. J. H.; Ramdin, M. Electrochemical Reduction of CO2 to Oxalic Acid: Experiments, Process Modeling, and Economics. *Ind. Eng. Chem. Res.* **2022**, *61* (40), 14837–14846. https://doi.org/10.1021/acs.iecr.2c02647.
(28) *Oxalic acid Price and Market Analysis - ECHEMI - ECHEMI.com*. ECHEMI. https://www.echemi.com/productsInformation/temppid160705011349-oxalic-acid.html (accessed 2023-11-08).
(29) *Caustic Soda Price per Ton April 2022 - News and Statistics - IndexBox*. https://www.indexbox.io/blog/caustic-soda-price-per-ton-april-2022/ (accessed 2023-11-08).
(30) *Caustic Soda Prices, Price, Pricing, Monitor | ChemAnalyst*. https://www.chemanalyst.com/Pricing-data/caustic-soda-3 (accessed 2023-11-08).
(31) *Sodium hydroxide Price List in Global Market - ECHEMI*. https://www.echemi.com/pip/caustic-soda-pearls-pd20150901041.html (accessed 2023-11-08).
(32) *Trisodium phosphate Price and Market Analysis - ECHEMI - ECHEMI.com*. ECHEMI. https://www.echemi.com/productsInformation/pd20161215175210037-trisodium-phosphate.html (accessed 2023-11-08).
(33) Jackson, E. *Improve Downstream Demand Accelerates the Price of Trisodium Phosphate in the Asian Market*. https://www.chemanalyst.com/NewsAndDeals/NewsDetails/improve-downstream-demand-accelerates-the-price-of-trisodium-phosphate-17297 (accessed 2023-11-08).
(34) *Nitric Acid Prices, News, Monitor | ChemAnalyst*. https://www.chemanalyst.com/Pricing-data/nitric-acid-1142 (accessed 2023-11-08).
(35) *Electricity data browser - Average retail price of electricity*. U.S. Energy Information Administration. https://www.eia.gov/electricity/data/browser/#/topic/7?agg=0,1&geo=0000001&endsec=2&linechart=ELEC.PRICE.FL-IND.M&columnchart=ELEC.PRICE.FL-IND.M&map=ELEC.PRICE.FL-IND.M&freq=M&ctype=columnchart&ltype=pin&rtype=s&maptype=0&rse=0&pin= (accessed 2023-11-08).

(36) Binnemans, K.; Jones, P. T. Rare Earths and the Balance Problem. *J. Sustain. Metall.* **2015**, *1* (1), 29–38. https://doi.org/10.1007/s40831-014-0005-1.
(37) Fritz, A. G.; Tarka, T. J.; Mauter, M. S. Technoeconomic Assessment of a Sequential Step-Leaching Process for Rare Earth Element Extraction from Acid Mine Drainage Precipitates. *ACS Sustainable Chem. Eng.* **2021**, *9* (28), 9308–9316. https://doi.org/10.1021/acssuschemeng.1c02069.
(38) *The "REE Basket Price" Deception And The Clarity Of Opex | Seeking Alpha*. https://seekingalpha.com/article/2996346-the-ree-basket-price-deception-and-the-clarity-of-opex (accessed 2025-01-21).
(39) Tsioka, M.; Voudrias, E. A. Comparison of Alternative Management Methods for Phosphogypsum Waste Using Life Cycle Analysis. *Journal of Cleaner Production* **2020**, *266*, 121386. https://doi.org/10.1016/j.jclepro.2020.121386.
(40) Wernet, G.; Bauer, C.; Steubing, B.; Reinhard, J.; Moreno-Ruiz, E.; Weidema, B. The Ecoinvent Database Version 3 (Part I): Overview and Methodology. *Int J Life Cycle Assess* **2016**, *21* (9), 1218–1230. https://doi.org/10.1007/s11367-016-1087-8.